\pdfoutput=1
\documentclass[bimj,fleqn]{w-art_edit}
\usepackage{amsmath}
\usepackage{times}
\usepackage{w-thm}
\usepackage[authoryear]{natbib}
\newcommand{\bcx}{{\bf X}}

\newcommand{\bcd}{{\bf D}}

\newcommand{\bck}{{\bf K}}
\newcommand{\bcp}{{\bf P}}
\newcommand{\bct}{{\bf T}}

\newcommand{\bx}{{\bf x}}

\newcommand{\bt}{{\bf t}}

\theoremstyle{plain}
\theoremstyle{definition}
\usepackage[]{graphicx}
\chardef\bslash=`\\ 

\usepackage{bm}
\usepackage[caption=false]{subfig}
\usepackage[flushleft]{threeparttable}
\DeclareMathOperator{\diag}{diag}
\usepackage[colorlinks=true, urlcolor=citecolor, linkcolor=citecolor, citecolor=citecolor]{hyperref}

\begin{document}
\pagespan{1}{}

\title[Flexible additive joint models]{Flexible Bayesian additive joint models with an application to type 1 diabetes research}
\author[Meike K\"ohler {\it{et al.}}]{Meike K\"ohler \footnote{Corresponding author: {\sf{e-mail: meike.koehler@helmholtz-muenchen.de}}, Phone: +49-(0)89-3068-2917, Fax: +49-(0)89-3187-3144}\inst{,1}} 
\address[\inst{1}]{Institute of Diabetes Research, Helmholtz Zentrum M\"unchen, and Forschergruppe Diabetes, Klinikum rechts der Isar, Technische Universit\"at M\"unchen, Germany}
\author[]{Nikolaus Umlauf \inst{2}}
\address[\inst{2}]{Department of Statistics, Faculty of Economics and Statistics,
Universit\"at Innsbruck, Austria}
\author[]{Andreas Beyerlein \inst{1}}
\author[]{Christiane Winkler \inst{1}}
\author[]{Anette-Gabriele Ziegler \inst{1, 3}}
\address[\inst{3}]{Forschergruppe Diabetes e.V. at the Helmholtz Zentrum M\"unchen, Germany}
\author[]{Sonja Greven \inst{4}}
\address[\inst{4}]{Department of Statistics, Ludwig-Maximilians-Universit\"at M\"unchen, Germany}

\begin{abstract}
The joint modeling of longitudinal and time-to-event data is an important tool of growing popularity to gain insights into the association between a biomarker and an event process. We develop a general framework of flexible additive joint models that allows the specification of a variety of effects, such as smooth nonlinear, time-varying and random effects, in the longitudinal and survival parts of the models. Our extensions are motivated by the investigation of the relationship between fluctuating disease-specific markers, in this case autoantibodies, and the progression to the autoimmune disease type 1 diabetes. By making use of Bayesian P-splines we are in particular able to capture highly nonlinear subject-specific marker trajectories as well as a time-varying association between the marker and the event process allowing new insights into disease progression. The model is estimated within a Bayesian framework and implemented in the R-package \texttt{bamlss}.
\end{abstract}

\maketitle                   
\thispagestyle{empty}

\section{Introduction}
\label{sec1}
\pagestyle{myheadings}
\renewcommand{\rightmark}{}

The joint modeling of longitudinal biomarkers and the time to disease onset or death offers unique insights into disease progression in various medical domains \citep{taylor_real-time_2013, gras_has_2013, daher_abdi_impact_2013}. Depending on the disease and the respective biomarker different challenges have to be faced in joint modeling. In the following, a general framework for the flexible joint modeling of longitudinal data and time-to-event is presented, which was motivated by unique cohort data from studies exploring the development of type 1 diabetes (T1D). The research on T1D underwent a paradigm shift in the past decade, when disease-specific autoantibodies  where shown to be diagnostic for the disease before the onset of clinical symptoms and thus paving the way for a pre-clinical diagnosis of T1D  \citep{ziegler_seroconversion_2013, bonifacio_predicting_2015, insel_staging_2015}. Prior to the onset of clinical symptoms, i.e. the need of insulin substitution, the disease is already progressing and insulin-producing beta-cells in the pancreas are gradually destroyed by the body's own immune system. This immune process, leading to an onset of clinical symptoms within months up to a decade, can be diagnosed by the emergence of T1D-specific autoantibodies. However, it remains an open question whether the longitudinal patterns of these autoantibodies might be associated with the rate of progression to T1D.

In recent years joint models gained larger popularity in the modeling of associations between time-varying biomarkers and time-to-event. By estimating a submodel for a longitudinal biomarker, usually a mixed model, jointly with the survival submodel of a time-to-event process, one can account for the informative censoring and the within-subject errors in the longitudinal model and can incorporate the longitudinal information, observed only at person-specific discrete timepoints, as a continuous-time covariate in the survival model. Comprehensive overviews on the topic are given in \cite{tsiatis_joint_2004}, \cite{rizopoulos_joint_2012} and \cite{gould_joint_2015}. In our work we focus on extensions of so-called shared parameter models. These assume that a set of parameters influences both the longitudinal and the survival model, and that there is conditional independence given those parameters. 

\begin{figure}[h] 
  \subfloat[\label{fig:long_obs}]{%
    \includegraphics[width=0.48\textwidth]{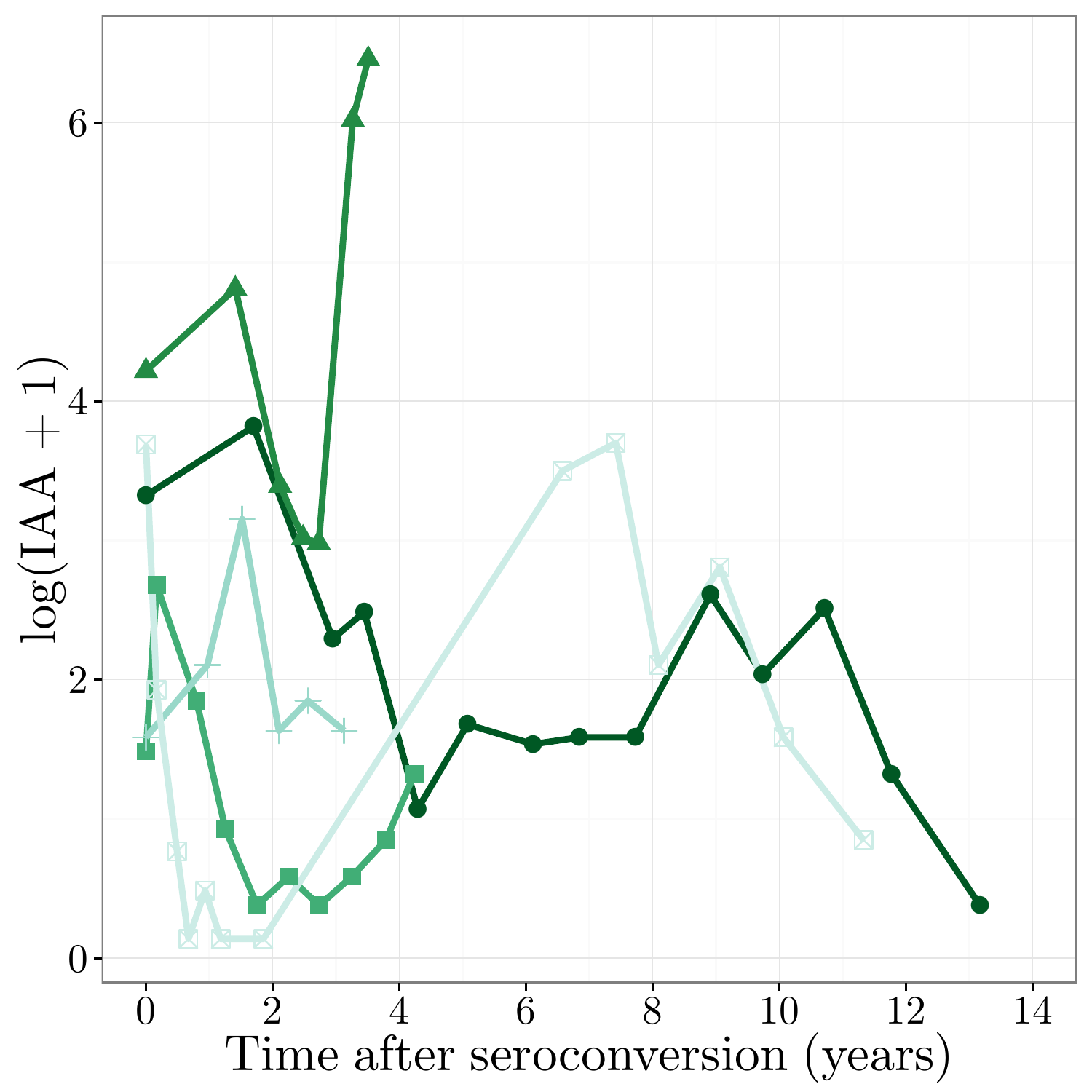} 
  } 
  \subfloat[\label{fig:long_est}]{%
    \includegraphics[width=0.48\textwidth]{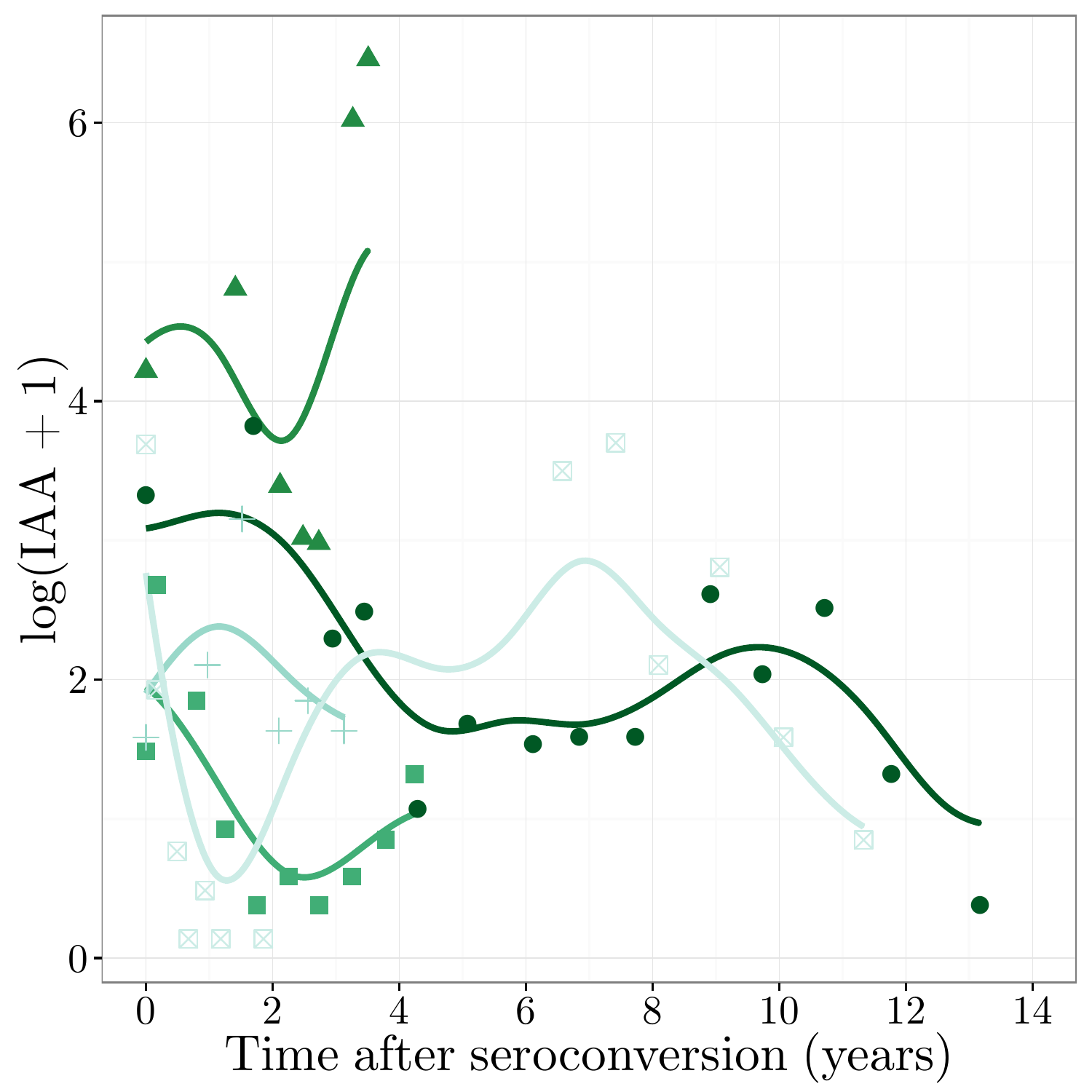} 
  } 
  \caption{Longitudinal marker values of $\log(IAA + 1)$ for five randomly selected subjects in the BABYDIAB/BABYDIET study. (a) Observed values (points) and linear interpolation (lines); (b) Observed values (points) and estimated trajectories (lines).} 
\label{fig:longitudinal}
\end{figure}

In T1D research little is known concerning typical trajectories of autoantibodies as biomarkers. At the same time the observed trajectories show highly nonlinear patterns over time and differ strongly between subjects, see Figure \ref{fig:long_obs}. In consequence, a flexible specification of individual trajectories in the longitudinal model is needed in our application.

 Much work on joint models has focused on simple parametric longitudinal trajectories, while only few approaches allow for more flexible, potentially non-parametric  longitudinal models. \cite{ding_modeling_2008-1} model mean trajectories by B-splines and allow for one multiplicative random effect per subject. For our application however it remains questionable if such a model is flexible enough to capture the highly different trajectories. Spline based approaches, that allow also the random effects to be non-linear functions in time, are mentioned by \cite{song_semiparametric_2008} and were employed by \cite{rizopoulos_bayesian_2011} and \cite{rizopoulos_combining_2014} as well as \cite{brown_flexible_2005} and \cite{brown_assessing_2009}.  While allowing for flexibility, a disadvantage of all these approaches is finding an optimal number of knots to specify the flexible longitudinal model, e.g. by AIC or DIC. As the number of random effects increases with the number of knots, this number is limited in practice. We aim to avoid the explicit choice of knots and number of basis functions by using a penalized spline approach, where a larger number of knots is specified and smoothness penalties are employed \citep{lang_bayesian_2004-1}. \cite{tang_semiparametric_2015} also make use of P-Splines in modeling longitudinal trajectories, but do so only in estimating the mean function, whereas we model also the individual trajectories as smooth functions of time. This is similar in spirit to the specification of individual trajectories in \cite{jiang_modelling_2015}, however we do not assume an underlying class membership for the random effects. 

The estimation of joint models with complex subject-specific trajectories poses a challenge to frequentist estimation approaches due to the necessary integration over potentially high-dimensional random effects distributions. Due to this drawback and further advantages of the Bayesian approach in joint modeling, such as straightforward model assessment and the potential integration of previous knowledge via priors \citep{gould_joint_2015}, many complex joint models, like e.g. the aforementioned models,  are specified within a Bayesian framework. 
The most widely used sampling approach for the parameter distributions in Bayesian joint models is Gibbs Sampling, e.g. \cite{faucett_simultaneously_1996, guo_separate_2004, brown_bayesian_2003}, also in conjunction with Metropolis-Hastings algorithms \citep{tang_semiparametric_2015}. In addition, the well established R-package \texttt{JMbayes} \citep{R_JMbayes, rizopoulos_r_2016} implementing \cite{rizopoulos_combining_2014} employs a random walk Metropolis-Hastings algorithm. Our Bayesian estimation approach is different as we employ a derivative-based Metropolis-Hastings algorithm, where we draw samples from approximations of the full conditionals using score vectors and Hessians of the parameters. Despite being computationally demanding this algorithm shows a high stability in the model estimation, as we also show in our simulations.

In addition to the need for a flexible longitudinal model, a further generalization of existing joint models seems necessary in our application, namely a time-varying association between the biomarker and the time-to-event. Here, the biomarker indicates an ongoing immune process eventually leading to the destruction of the insulin-producing beta cells. As the activity of the immune system is constantly regulated, it is plausible that the association between a biomarker and the hazard of T1D varies over time. For example a recent paper by \citet{meyer_aire-deficient_2016} indicated that patients with an autoimmune disease can also present unique disease-ameliorating autoantibodies. Such a time-varying association has rarely been studied in the context of joint models. Using a discretized time-scale and a probit model for the discrete hazard function, \cite{barrett_joint_2015} allow for the association to vary over the discrete time points in their model. However this flexible specification is not considered in their simulations, the applied examples or the code provided to fit the models. A time-varying coefficient to associate the marker and the event process is the focus of the conditional score estimation approach in \cite{song_semiparametric_2008}. This approach can be seen as a weighted local partial likelihood without any assumptions on the distribution of the random effects. While this approach accounts for measurement error and short-term biological fluctuations in the longitudinal marker when modeling the hazard, it only permits inference on the survival parameters and not on the longitudinal model.

In order to allow for these two extensions, the flexible longitudinal trajectories and a potentially nonlinear time-varying association, both modeled by penalized splines, we develop and implement a highly flexible framework for joint models available within the R-package \texttt{bamlss}\nocite{R_bamlss}. As we represent all parts of this flexible joint model as structured additive predictors, which can include linear, parametric but also nonparametric penalized terms, we are able to allow potentially nonlinear, smooth, random, and time-varying effects in both submodels. In consequence the possibilities of this implementation go way beyond the two extensions that originally  triggered the development. By applying this flexible model to the combined data set from two German high-risk T1D birth cohorts we aim to shed further light on the complex relationship between T1D-associated autoantibodies and the onset of clinical disease. 

The remainder of this paper is structured as follows: The general model structure and potential extensions are outlined in Section \ref{sec2}. In Section \ref{sec3}, details on the Bayesian estimation procedure are given. A thorough testing of the model estimation through simulations is presented in Section \ref{sec4} and the application to our T1D research question in Section \ref{sec5}. Concluding remarks are given in Section \ref{sec6} and technical details as well as additional figures can be found in the Appendix. The presented model is implemented in the R-package \texttt{bamlss} \citep{R_bamlss}. Source code to reproduce the simulation results is available in the ancillary material. 

\section{Methods}
\label{sec2}
In the following, the general setup for additive joint models is presented with a special focus on two extensions in the present work compared to existing approaches: the flexible specification of longitudinal trajectories as well as the time-varying association between longitudinal marker and event. An overview of potential further model specifications illustrates the flexibility of the presented model family.

\subsection{General Setup}
For every subject $i=1,\ldots, n$ we observe a potentially right-censored follow-up time $T_i$ and the event indicator $\delta_{i}$ (1 if subject $i$ experiences the event, 0 if it is censored). We model the hazard of an event at time $t$ as 
\begin{equation}
h_{i}(t)= \exp \left\{\eta_i(t)\right\} = \exp\left\{\eta_{\lambda i}(t)+
\eta_{\gamma i}+\eta_{\alpha i}(t) \cdot \eta_{\mu i}(t) \right\}
\label{eq:hazard}
\end{equation}

including in the full predictor $\eta$ a predictor $\eta_{\lambda}$  for all survival covariates that are time-varying or have a time-varying coefficient including the log baseline hazard, a predictor for baseline survival covariates $\eta_{\gamma}$ as well as a predictor $\eta_{\alpha}$ representing the potentially time-varying association between the longitudinal marker $\eta_{\mu}$ and the hazard. \\
We also observe a longitudinal response $\bm{y}_{i}=[y_{i1}, \cdots, y_{i n_i}]^{\top}$ at the potentially subject-specific ordered time points $\bt_i=[t_{i1}, \cdots, t_{i n_i}]^{\top}$  with $t_{i1}\leq \cdots \leq t_{in_{i}}\leq T_{i}$. $\bt=[\bt^{\top}_1, \cdots, \bt^{\top}_n]^{\top}$ denotes the vector of the $N = \sum_{i=1}^{n} n_i$ longitudinal measurement time points of all subjects. The longitudinal response at $t_{ij}$ with $j=1,\ldots,n_{i}$ is modeled as
\begin{equation}
y_{ij}=\eta_{\mu i}(t_{ij})+\varepsilon_{ij}
\label{eq:mu}
\end{equation}
with independent errors $\varepsilon_{ij} \sim N(0, \exp[\eta_{\sigma i}(t_{ij})]^2)$ allowing to also model the error variance. Thus $\eta_{\mu i}(t_{ij})$ represents the longitudinally observed marker value without error at timepoint $t_{ij}$. This \textquotedblleft true\textquotedblright  $\:$ marker value serves as a continuous-time covariate in the hazard in \eqref{eq:hazard} and links the two model equations. \\

Each predictor $\eta_{ki}$ with $k \: \in \: \{\lambda, \gamma, \alpha, \mu, \sigma \}$ is a structured additive predictor, i.e. a sum of $M_k$ functions of covariates $\bm{x}_i$,
\begin{equation*}
\eta_{ki} =  \sum_{m=1}^{M_k} f_{km}(\bm{x}_{ki}).
\end{equation*} 
Different subsets $\bm{x}_{ki}$ of $\bm{x}_i$ can serve as covariates for the different predictors, with each $f_{km}$ typically depending on one or two covariates. For time-varying predictors the functions can also dependent on time $\eta_{ki}(t) =  \sum_{m=1}^{M_k} f_{km}(\bm{x}_{ki}(t), t)$ with a potentially time-varying covariate vector $\bm{x}_{ki}$. 
We express the vector of predictors for all subjects as $\bm{\eta}_k = [\eta_{k1}, \cdots, \eta_{kn}]^{\top}$. These vectors are of length $n$ for the survival part of the model \eqref{eq:hazard}, where $\bm{\eta}_k(t)$ for $k \in \{\lambda, \alpha, \mu\}$ denotes that predictors are evaluated at time $t$. In the longitudinal part of the model \eqref{eq:mu} the vector $\bm{\eta}_k(\bt)$ for  $k \in \{\mu, \sigma \}$ is of length $N$, containing entries $\eta_{ki}(t_{ij})$ for all $j=1,\ldots, n_i$, $i=1,\ldots, n$ , i.e. evaluations at all observed time points $\bt$ for the corresponding subjects.

The functions $f_{km}(\bm{x}_{ki})$ can model a variety of effects, such as smooth, spatial, time-varying or random effects terms which can be expressed in a straightforward notation for every term $m$ of predictor $k$ by using suitable basis function expansions and corresponding penalties $\bcp_{km}$. In a generic setup we let

\begin{equation}
\bm{f}_{km} = \bcx_{km} \bm{\beta}_{km} \text{\qquad and \qquad} \bcp_{km}=\frac{1}{\tau_{km}^2} \bm{\beta}_{km}^{\top} \bck_{km} \bm{\beta}_{km},
\label{eq:design_penalty}
\end{equation}
with the vector of function evaluations $\bm{f}_{km}$ stacked for each subject, the design matrix $\bcx_{km}$, the coefficient vector $\bm{\beta}_{km}$, the penalty matrix  $\bck_{km}$ and the variance parameter $\tau_{km}^2$ that controls the amount of penalization of the respective term. In the Bayesian setting a penalization is imposed by specifying an appropriate prior distribution for the parameters, $\bm{\beta}_{km} \sim N(\bm{0}, [\frac{1}{\tau_{km}^2} \bck_{km} ]^{-})$ with $\bm{A}^{-}$ denoting the generalized inverse of $\bm{A}$, as presented in more detail in section \ref{est_bayes}. Note that these basic penalties can be extended further as shown in more detail in the next subsection. 

In analogy to the differences in form in the generic vector of predictors, i.e. $\bm{\eta}_k$, $\bm{\eta}_k(t)$ and $\bm{\eta}_k(\bt)$, the form of the generic vectors of function evaluations $\bm{f}_{km}$ and the generic design matrices $\bcx_{km}$ also differs between predictors and submodels. For ease of notation we drop the subscript $m$ for the different terms per predictor in this illustration. For the time-constant survival predictor $\bm{\eta}_{\gamma}$, we observe a vector of covariates $\bm{x}_{\gamma i}$ for every subject and stack these in the design matrix $\bcx_{\gamma}=[\bm{x}_{\gamma 1}, \cdots, \bm{x}_{\gamma n}]^{\top}$ of size $n \times p_{\gamma}$ resulting in the vector of function evaluations $\bm{f}_{\gamma}=[f_{\gamma}(\bm{x}_{\gamma 1}), \cdots, f_{\gamma}(\bm{x}_{\gamma n})]^{\top}$. For the time-varying predictors of the survival part, i.e. $k \in \{\lambda, \alpha, \mu\}$,  $\bm{x}_{k i}(t)$ denotes the subject covariate vector, including basis evaluations for non-linear effects over time at time $t$, resulting in the design matrix $\bcx_k(t)$ of evaluations of size $n \times p_{k}$ and the vector $\bm{f}_{k}(t)=[f_{k}(\bm{x}_{k 1}(t), t), \cdots, f_{k}(\bm{x}_{k n}(t), t)]^{\top}$ of length $n$ for each $t$. Finally for predictors in the longitudinal submodel, i.e. $k \: \in \: \{\mu, \sigma\}$, we observe the $n_i \times p_k$ covariate matrix $\bm{x}_{k i}(\bt_i)$ for every subject $i$ at the $n_i$ subject-specific time points, resulting in the $N \times p_k$ stacked design matrix $\bcx_{k}(\bt)$ for all subjects at all timepoints with the vector $\bm{f}_{k}(\bt)=[f_{k}(\bm{x}_{k 1}(\bt_1), \bt_1)^{\top}, \cdots, f_{k}(\bm{x}_{k n}(\bt_n), \bt_1)^{\top}]^{\top}$.  \\

To illustrate how different effects are subsumed under this notation by the specification of the respective design matrices, we formulate a standard shared parameter joint model within this framework. Note that we drop the index $m$ for predictors which consist of only one term. We specify the log-baseline hazard $\bm{\eta}_{\lambda}(t)=\bm{f}_{\lambda}(t)=\bcx_{\lambda}(t)\bm{\beta}_{\lambda}$ as a smooth function in time by P-splines with a B-spline basis, $f_{\lambda}(t)= \sum_{d=1}^{D} \beta_d B_d(t)=: \bm{x}^{\top}_{\lambda}(t) \bm{\beta}_{\lambda}$, and the penalty matrix $\bck_{\lambda}= \bcd_r^\top\bcd_r$ with $\bcd_r$ as the $r$-th difference matrix of appropriate dimension \citep{EilersMarx1996}. In the Bayesian setting, using Bayesian P-Splines, smoothing is induced by appropriate prior specification, where the difference penalties are replaced by their stochastic analogues, i.e. random walks \citep{lang_bayesian_2004-1}.  Here $\bm{x}_{\lambda}(t)$  contains the evaluation of the $D$ B-spline basis functions $B_d(t)$ at time $t$. As the baseline hazard is not subject-specific, $\bcx_{\lambda}(t)$ contains $n$ stacked  replications of $\bx_{\lambda}(t)$. Parametric effects of baseline survival covariates are modeled as $\bm{\eta}_{\gamma}=\bm{f}_{\gamma} = \bcx_{\gamma}\bm{\beta}_{\gamma}$, where each row of $\bcx_{\gamma}$ contains the subject-specific covariate vector and $\bck_{\gamma}$ is taken as the zero-matrix $\bm{0}$. The usual time-constant association between longitudinal and survival model is implemented as $\bm{\eta}_{\alpha}=\bm{f}_{\alpha} = \bm{1}_n \beta_{\alpha}$, where $\bm{1}_n$ is a vector of ones of length $n$ and $\bck_{\alpha}=\bm{0}$. The predictor vector for the longitudinal part with a random intercept in the linear mixed effects model can be specified as $\bm{\eta}_{\mu}(\bt)= \bm{f}_{\mu 1}(\bm{t}) + \bm{f}_{\mu 2}(\bm{t}) =  \bcx_{\mu 1}(\bm{t})\bm{\beta}_{\mu 1} + \bcx_{\mu 2}(\bm{t})\bm{\beta}_{\mu 2}$, where $\bcx_{\mu 1}(\bm{t})\bm{\beta}_{\mu 1}$ are the design matrix and coefficient vector of the fixed effects potentially including a parametric effect of time with $\bck_{\mu 1}=\bm{0}$, and $\bcx_{\mu 2}(\bm{t})\bm{\beta}_{\mu 2}$ is a basis matrix representation of a random intercept. In more detail $\bcx_{\mu 2}$ is an $N \times n$ indicator matrix, where the $i$th column indicates which longitudinal measurements belong to subject $i$, $\bm{\beta}_{\mu 2} = [ \beta_{\mu 2 1}, \cdots, \beta_{\mu 2 n} ]$ denotes the coefficient vector and an $n \times n$ identity matrix as penalty $\bck_{\mu 2}=\bm{I}_n$ ensures $\beta_{\mu 2 i} \sim N(0, \tau^2_{\mu 2} )$ independently. Finally the error variance is modeled as constant using $\bm{\eta}_{\sigma}(\bt) = \bm{f}_{\sigma}(\bt) =  \bm{1}_N \beta_{\sigma}$ with $\bck_{\sigma}=\bm{0}$.


\subsection{Important extensions of current models}
A special focus in our joint model approach lies on the flexibility of the longitudinal predictor $\eta_\mu$. We model the trajectory for every subject as the sum of fixed covariate effects, a smooth function of time, a random intercept as well as  smooth subject-specific deviations from this function over time,
\begin{equation}
\eta_{\mu i}(t)= f_{\mu 1}\left(t\right) + f_{\mu 2}\left(i\right) +  f_{\mu 3}\left(t, i\right) + \sum_{m=4}^{M_{\mu}} f_{\mu m}\left(\bm{x}_{\mu m}\right). 
\label{eq:fct_int}
\end{equation}

In this parameterization $f_{\mu 1}(t)$ is a smooth effect of time, constructed like $f_{\lambda}(t)$, and $f_{\mu 2}(i)$ is a random intercept as illustrated in the previous sub-section. The term $f_{\mu 3}(t, i)$ denotes the smooth subject-specific deviations from the global time effect using functional random intercepts \citep{scheipl_functional_2015}. Additionally linear or parametric effects, including a global intercept, as well as further smooth effects of covariates can be represented by an extra term in $\sum_{m=4}^{M_{\mu}} f_{\mu m}(\bm{x}_{\mu m})$. 
The basis for the functional random intercepts can be specified within the basis function approach as row tensor products of the marginal basis of a random intercept, marked by the subscript $s$, and the marginal basis for a smooth effect of time, marked by the subscript $t$. We denote the vector of function evaluations at every observed longitudinal time point in $\bt$ for the corresponding subjects in  $\bm{i}= [1, \cdots, n]^{\top}$  as 
\begin{align}
f_{\mu 3}\left(\bt, \bm{i}\right) = (\bcx_{\mu 3 s} \odot \bcx_{\mu 3 t}) \bm{\beta}_{\mu 3} = \bcx_{\mu 3} \bm{\beta}_{\mu 3},
\end{align}
where $\bcx_{\mu 3 s}$ is an $N \times n$ indicator matrix as the basis for a random intercept as specified for $\bcx_{\mu 2}$ in the previous sub-section, $\bcx_{\mu 3 t}$ is an $N \times D$ matrix of evaluations of a marginal spline basis at $\bt$ and $\bcx_{\mu 3}$ is the $N \times nD$ basis matrix resulting from the row tensor product. The row tensor product $\odot$ of a $p \times a$ matrix $\bm{A}$ and a $p \times b$ matrix $\bm{B}$ is defined as the $p \times ab$ matrix $\bm{A} \odot \bm{B}=(\bm{A} \otimes \bm{1}_b^\top ) \cdot (\bm{1}_a^\top \otimes \bm{B})$ with $\cdot$ denoting element-wise multiplication.

The corresponding penalty term is constructed from the marginal penalty matrices: 
\begin{equation}
\bcp_{\mu 3} = \bm{\beta}_{\mu 3}^\top \left( \frac{1}{\tau_{\mu 3 s}^2} \bck_{\mu 3 s} \otimes \bm{I}_t + \frac{1}{\tau_{\mu 3 t}^2} \bm{I}_{s} \otimes \bck_{\mu 3 t} \right) \bm{\beta}_{\mu 3} = \bm{\beta}_{\mu 3}^\top \left( \frac{1}{\tau_{\mu 3 s}^2} \bm{\tilde{K}}_{\mu 3 s} + \frac{1}{\tau_{\mu 3 t}^2}  \bm{\tilde{K}}_{\mu 3 t} \right) \bm{\beta}_{\mu 3},
\label{eq:aniso_pen}
\end{equation}
where $\otimes$ denotes the Kronecker product, $\bck_{\mu 3 s}=\bm{I}_n$ is the penalty matrix for the random effect and $\bck_{\mu 3 t}$ is an appropriate penalty matrix for the smooth effect of time such as a difference penalty for B-splines. The enlarged penalty matrices $\bm{\tilde{K}}_{\mu 3 s}$ and $\bm{\tilde{K}}_{\mu 3 t}$ yield a penalization for every subject, resulting in a random effects structure and a smoothness penalization across time for each subject. Note that by specifying two variance parameters, $\tau_{\mu 3 s}^2$ and $\tau_{\mu 3 t}^2$, the amount of penalization can differ in the direction of time and across subjects, resulting in an anisotropic penalty. This specification allows for a highly flexible modeling of individual trajectories over time. 

Given the specification of a separate global intercept and subject-specific random intercepts, the constraints $\int f_{\mu 1}(t) dt=0$ and $\int f_{\mu 3}(t, i) dt=0$ for every $i$ are set in order to ensure identifiability. The necessary linear constraint $\int f_{\mu 1}(t) dt=0$ is implemented for B-splines by transforming the marginal basis $\bcx_{\mu 3 t}$ into an $ N \times (D-1)$ matrix $\bm{\dot{X}}_{\mu 3 t}$ for which it holds that $\bm{\dot{X}}_{\mu 3 t} \bm{1}_{D-1}= \bm{0}$ as shown in \citet[][chapter 1.8]{wood_generalized_2006}, and adjusting the penalty accordingly. Transforming the marginal basis and constructing the row tensor product $\bcx_{\mu 3}$ using the transformed basis matrix $\bm{\dot{X}}_{\mu 3 t}$ with correspondingly adjusted marginal penalty ensures that the identification constraint $\int f_{\mu 3}(t, i) dt=0$ for every $i$ is also fulfilled.\\

As a second extension to existing shared-parameter models we also specify the association between longitudinal and survival model as a structured additive predictor $\eta_{\alpha}$. In consequence, this predictor can be modeled as a function of time and/or other covariates. 
Motivated by our applied research questions we model $\eta_{\alpha}(t)= f_{\alpha}(t)$ as a smooth function of time by using penalized splines, as specified for the baseline hazard. This allows us to find patterns beyond the standard joint model specification to explain the relationship between longitudinal marker and survival process. These patterns could for example be critical time windows in which a non-zero effect of $\eta_{\alpha}$ is present or a potential change in the direction of the association $\eta_{\alpha}$ over time.

\subsection{Further potential specifications} 
The presented general framework of structured additive joint models allows for a variety of different effect specifications by making use of the flexibility of Bayesian structured additive regression models \citep{fahrmeir_penalized_2004} as well as adding functional extensions \citep{scheipl_functional_2015}. Besides the presented smooth, time-varying, random effects and functional random intercept terms, a variety of further effects can be incorporated. 
Table \ref{tbl:effects} gives an overview of possible terms. All these terms can be specified by formulating the desired effect in a basis function representation with an appropriate penalty term. 

\begin{table}[h]
\begin{center}
\caption{Effects $f_{km}(\bm{x}_{ki})$ that can be specified within a predictor $\eta_{k}$ in structured additive joint models; modified from a similar table in  \cite{scheipl_functional_2015}.}
\label{tbl:effects}
\begin{tabular}{lll}
\hline
covariate (subset of $\bm{x}$) & $f_{km}(\bm{x}_{k})$ constant over $t$ & $f_{km}(\bm{x}_{k})$ varying over $t$\\
\hline
no covariate         & scalar intercept $1 \cdot \beta$  & smooth effect of time $f(t)$ \\
scalar covariate $z$ & linear effect $z  \cdot \beta$    & linear effect varying over time $z  \cdot f(t)$\\
                     & smooth effect $f(z)$    & smooth effect over time $f(z,t)$\\
spatial covariate(s) $s$ & spatial effect $f(s)$  & spatial effect over time $f(s,t)$\\

grouping variable $g$  & random intercept $\beta_g$ & functional random intercept $f_{g}(t)$\\
scalar and grouping variable & random slope $z \cdot \beta_g$ & functional random slope $z \cdot f_g(t)$\\
vector of scalars $[z_1, z_2]$ & linear interaction $z_1 \cdot z_2 \cdot \beta$  & linear interaction over time $z_1 \cdot z_2 \cdot f(t)$  \\
                               & varying coefficient $z_1 \cdot f(z_2)$  & \\ 
                               & smooth effect $f(z_1, z_2)$  &  \\
\hline
\end{tabular}
\end{center}
\end{table}

For details on the specification of such effects please refer to \cite{fahrmeir_penalized_2004, scheipl_functional_2015, wood_generalized_2006}. Further details on the practical aspects within our implementation are given in section \ref{est_implementation}.

\section{Estimation}
\label{sec3}

We estimate the model in a Bayesian framework using  Newton-Raphson and Markov chain Monte Carlo (MCMC) algorithms.

\subsection{Likelihood}
\label{est_likelihood}
Under the assumption of conditional independence of the survival outcomes $[T_i, \delta_i]$ and the longitudinal outcome $\bm{y}_i$, given the random effects, the likelihood of the specified joint model is the product of the two submodel likelihoods $L^{surv}$ and $L^{long}$ for the survival and the longitudinal model 

\begin{equation*}
L\left[\bm{\theta} | \bct, \bm{\delta}, \bm{y}\right] = L^{surv}\left[\bm{\eta}_{\lambda}(\bct),
\bm{\eta}_{\gamma},\bm{\eta}_{\alpha}(\bct),\bm{\eta}_{\mu}(\bct)\right] \cdot
 L^{long}\left[\bm{\eta}_{\mu}(\bt),\bm{\eta}_{\sigma}(\bt)\right],   
\end{equation*}
where $\bm{\theta}$ is the vector of all parameters in the model and $\bct=[T_1, \cdots T_n]^{\top}$, $\bm{\delta}=[\delta_1, \cdots \delta_n]^{\top}$, and $\bm{y}=[\bm{y}^{\top}_1, \cdots \bm{y}_n^{\top}]^{\top}$ are the response vectors. The additive predictors implicitely also depend on covariates and model parameters. The log-likelihood of the survival part is
\begin{equation}
\ell^{surv}\left[\bm{\eta}_{\lambda}(\bct),\bm{\eta}_{\gamma},\bm{\eta}_{\alpha}\left(\bct\right),\bm{\eta}_{\mu}\left(\bct\right)\right] =  \bm{\delta}^{\top}\bm{\eta}(\bct)-\bm{1}_{n}^{\top}\bm{\Lambda}\left(\bct\right),
\label{eq:l_surv}
\end{equation}
where $\bm{\Lambda}(\bct)=[\Lambda_{1}(T_{1}),\ldots,\Lambda_{n}(T_{n})]^{\top}$ is the vector of the cumulative hazard rates \\
$\Lambda_{i}(T_{i})=
\exp(\eta_{\gamma i})\int_{0}^{T_{i}}\exp[\eta_{\lambda i}(u)+\eta_{\alpha i}(u)\cdot \eta_{\mu i}(u)]du$ and $\bm{\eta}(\bct)=[\eta_1(T_1), \cdots, \eta_n(T_n) ]$ denotes the vector of the full predictors evaluated at the subject-specific survival times. The log-likelihood of the longitudinal part of the model is
\begin{equation}
\ell^{long}\left[\bm{\eta}_{\mu}\left(\bt\right),\bm{\eta}_{\sigma}\left(\bt\right)\right] = -\frac{N}{2}\log(2\pi)- \bm{1}_{N}^{\top} \bm{\eta}_{\sigma}\left(\bt\right) -\frac{1}{2}(\bm{y}-\bm{\eta}_{\mu}\left(\bt\right))^{\top} \bm{R}^{-1}(\bm{y}-\bm{\eta}_{\mu}\left(\bt\right)).
\label{eq:l_long}
\end{equation}
$\bm{\eta}_{\mu}(\bt)$ and $\bm{\eta}_{\sigma}(\bt)$ are the predictor vectors of length $N$ corresponding to the longitudinal response $\bm{y}$ and $\bm{R}=\text{blockdiag} (\bm{R}_{1}, \cdots, \bm{R}_{n})$, where $\bm{R}_{i}$ can reflect the error structure of interest. In our case, we assume $\bm{R}_{i}=\diag (\exp[\eta_{\sigma i}(t_{i1})]^2, \cdots, \exp[\eta_{\sigma i}(t_{in_i})]^2)$ so that $\bm{R}$ reduces to a diagonal matrix.

\subsection{Priors and Posterior}
\label{est_posterior}
In this general framework above, a variety of terms (cf. Table \ref{tbl:effects}) can be specified using corresponding priors. For linear or parametric terms we use vague normal priors on the vectors of the regression coefficients, e.g. $\bm{\beta}_{km} \sim N(0, 1000^2)$, approximately corresponding to the precision matrices $\bck_{km}=\bm{0}$ as explained above. Smooth and random effect terms are regularized by placing suitable multivariate normal priors on the coefficients
\begin{equation*}
p(\bm{\beta}_{km}|\tau_{km}^2) \propto \left(\frac{1}{\tau_{km}^2}\right)^{\frac{\operatorname{rank}\left(\bck_{km}\right)}{2}} \exp \left(\frac{1}{2\tau_{km}^2} \bm{\beta}_{km}^\top \bck_{km} \bm{\beta}_{km} \right)
\end{equation*}
with precision matrix $\bck_{km}$ as specified in the penalty \eqref{eq:design_penalty}. We use independent inverse Gamma hyperpriors $\tau_{km}^2 \sim IG(0.001, 0.001)$ to obtain an inverse Gamma full conditional for the variance parameters. In addition to the inverse gamma distribution, different priors are possible for the variance parameters in our implementation, such aus Half-Cauchy and Half-normal distributions. The variance parameters $\tau^2_{km}$ control the trade-off between flexibility and smoothness in the nonlinear modeling of effects. As such they can be interpreted analogous to inverse smoothing parameters in a frequentist approach. 

For anisotropic smooths, when multiple variance parameters $\bm{\tau}_{km}^2 = (\tau_{kms}^2, \tau_{kmt}^2)$ are involved as in \eqref{eq:aniso_pen}, we use the prior 
\begin{equation}
p(\bm{\beta}_{km}|\bm{\tau}_{km}^2) \propto \left|\frac{1}{\tau_{kms}^2}    \bm{\tilde{K}}_{kms}   + \frac{1}{\tau_{kmt}^2}  \bm{\tilde{K}}_{kmt}  \right|^{\frac{1}{2}}
\exp \left( -\frac{1}{2} \boldsymbol{\beta}_{km}^\top \left[\frac{1}{\tau_{kms}^2}    \bm{\tilde{K}}_{kms}  + \frac{1}{\tau_{kmt}^2}  \bm{\tilde{K}}_{kmt}  \right] \boldsymbol{\beta}_{km}  \right).
\label{eq:aniso_prior}
\end{equation}

The resulting posterior of the model is
\begin{equation*}
\begin{split}
p(\bm{\theta} | \bct, \bt,  \bm{\delta}, \bm{y}) \propto L^{surv}\left[\bm{\eta}_{\lambda}(\bct),
\bm{\eta}_{\gamma},\bm{\eta}_{\alpha}(\bct),\bm{\eta}_{\mu}(\bct)\right] \cdot
 L^{long}\left[\bm{\eta}_{\mu}(\bt),\bm{\eta}_{\sigma}(\bt)\right]  \\
\cdot \prod_{k \in \{\lambda, \gamma, \alpha, \mu, \sigma\} } \prod_{m=1}^{M_k} \left[p(\bm{\beta}_{km}|\bm{\tau}_{km}^2)  p(\bm{\tau}_{km}^2)\right],
\end{split}
\end{equation*}

where $p(\bm{\beta}_{km}|\bm{\tau}_{km}^2)$ are the priors of the vectors of regression parameters and $p(\bm{\tau}_{km}^2)$ are the priors of the variance parameters for each term $m$ and predictor $k$.

\subsection{Bayesian Estimation}
\label{est_bayes}

Point estimates of $\bm{\theta}$ can be obtained by posterior mode and posterior mean estimation. We estimate the posterior mode by maximizing the log-posterior of the model using a Newton-Raphson procedure, the posterior mean is obtained via derivative-based Metropolis-Hastings sampling and thus computationally demanding. We therefore recommend to use posterior mode estimates for a first quick assessment of the model and in order to obtain starting values for the posterior mean sampling. 

In the maximization of the log-posterior to obtain the posterior mode, we update blockwise each term $m$ of predictor $k$ in each iteration $l$ as
\[
\bm{\beta}_{km}^{[l+1]}=\bm{\beta}^{[l]}_{km}- \nu^{[l]}_{km} \bm{H}\left(\bm{\beta}_{km}^{[l]}\right)^{-1}\bm{s}\left(\bm{\beta}_{km}^{[l]}\right)
\]
with potentially varying steplength $\nu^{[l]}_{km}$ and with the score vector $\bm{s}(\bm{\beta}_{km} )$ and the Hessian $\bm{H}(\bm{\beta}_{km} )$, which can be found in Appendix A. We optimize the variance parameters in each updating step to minimize the corrected AIC \citep[AICc,][]{hurvich_smoothing_1998}, which showed good performance in smoothing parameter estimation in \citet{belitz_simultaneous_2008}. Additionally we optimize the steplength $\nu_{km}^{[l]}$ over $(0, 1]$ in each step to maximize the log-posterior. We assume the coefficients to have an approximately normal posterior distribution and derive credibility intervals from $N(\hat{\bm{\beta}}_{km}, [-\bm{H}(\hat{\bm{\beta}}_{km}) ]^{-1} )$ for quick approximate inference. 

For the posterior mean sampling we construct approximate full conditionals $\pi(\bm{\beta}_{km}| \cdot)$ based on a second order Taylor expansion of the log-posterior centered at the last state $\beta_{km}^{[l]}$, similar to \cite{fahrmeir_penalized_2004}, \cite{Klein+Kneib+Klasen+Lang:2015} and \cite{Klein+Kneib+Lang+Sohn:2015}. 
The proposal density from this approximate full conditional is proportional to a multivariate normal distribution with the precision matrix $(\bm{\Sigma}^{[l]}_{km})^{-1} = -\bm{H}(\bm{\beta}^{[l]}_{km})$ and the mean $\bm{\mu}^{[l]}_{km} = \bm{\beta}^{[l]}_{km} - \bm{H}(\bm{\beta}^{[l]}_{km})^{-1} \bm{s}(\bm{\beta}_{km}^{[l]})$.
In each iteration $l$ of the sampler and for updating block $km$ a candidate $\bm{\beta}_{km}^\ast$ is drawn from the proposal density

\begin{align*}
q(\bm{\beta}^\ast_{km} | \bm{\beta}^{[l]}_{km}) = N(\bm{\mu}_{km}^{[l]},\bm{\Sigma}^{[l]}_{km})
\end{align*}
and is accepted with the probability
 \begin{align*}
a(\bm{\beta}^\ast_{km} | \bm{\beta}^{[l]}_{km}) = \min \left( 
\frac{\pi(\bm{\beta}^\ast_{km} | \cdot ) q(\bm{\beta}^{[l]}_{km} | \bm{\beta}^\ast_{km})}
{\pi(\bm{\beta}^{[l]}_{km} | \cdot ) q(\bm{\beta}^\ast_{km} | \bm{\beta}^{[l]}_{km})}, 1\right),
\end{align*}
where $\pi(\bm{\beta}^\ast_{km} | \cdot )$ is the full conditional for the candidate and $\pi(\bm{\beta}^{[l]}_{km} | \cdot)$ is the full conditional for the current iterate. By drawing candidates from a close approximation of the full conditional, using the log-posterior centered at the previous state, we approximate a Gibbs sampler and achieve high acceptance rates and good mixing.

For the sampling of the variance parameters $\tau^2_{km}$ Gibbs sampling is employed, as the full conditionals $\pi(\tau^2_{km}|\cdot)$ follow an inverse Gamma distribution, if inverse Gamma hyperpriors are used. Slice sampling is employed when no simple closed-form full conditional can be obtained as for example in the sampling of variance parameters for anisotropic smooths \eqref{eq:aniso_prior} or for other hyperpriors.

\subsection{Implementation details}
\label{est_implementation}
The model estimation is implemented within R \citep{R} in the package \texttt{bamlss} \citep{R_bamlss} that allows the Bayesian estimation of a variety of models within the framework of Bayesian Additive Models for Location, Scale and Shape. The specification of appropriate design matrices and penalties for the desired effects is conducted internally via the R-package \texttt{mgcv} \citep{Wood_fast_2011}. In consequence the full range of implemented smoothing approaches, such as P-splines, thin-splate regression splines, random effects, and Markov Random Fields, can be used within our implementation. We refer to \cite{wood_generalized_2006} and \cite{WoodPyaSaefken2016}  for further information on  model terms, bases and penalities. In our model specification in the simulations and the application we make use of Bayesian P-splines \citep{lang_bayesian_2004-1} to model smooth effects.
As the integrals in the survival likelihood as well as in the respective scores and Hessians have no analytical solution, they are approximated numerically using the trapezoidal rule and a fixed number of 25 integration points. 
Starting values for the posterior mean sampling are obtained by estimating the posterior mode of the joint model. 
The posterior mean sampling is implememented to potentially run in parallel on a number of specified cores on Linux systems. More details can be found in the documentation of the \texttt{bamlss} R-package.

\section{Simulation}
\label{sec4}
We assess the estimation of our model by means of a simulation study with focus on two aspects: First, comparing our results with the established joint model implementation in \texttt{JMbayes} \citep{R_JMbayes} for models with time-constant $\eta_{\alpha}$. Second, we want to assess the ability to model highly complex longitudinal trajectories as well as a time-varying effect of $\eta_{\alpha}(t)$, the two important new extensions within our framework. With this simulation we also aim to gain insights into the estimation quality of the model when applied to real data sets of T1D cohorts that motivated our methods development. Therefore we simulate two differing data situations, mimicking real cohort data. The first simulated data setting, corresponding to the cohort data presented in the Application Section, has less subjects, at more variably spaced time points but with a longer follow up, than the other. Finally we aim to assess how well the posterior mode estimation can approximate the effects in comparison with the posterior mean estimates.

\subsection{Simulation design}
For every setting we generate longitudinal measurements for $n$ subjects at a fixed grid of time points $\mathcal{P}$ based on a true longitudinal model $\eta_{\mu}(t)$ as specified in \eqref{eq:fct_int} with the time effect $ f_{\mu 1}\left(t\right) = 0.1(t+2)\exp(-0.075t)$, the random intercepts $f_{\mu 2}\left(i\right) = r_{i}$ where $r_{i} \sim N(0, 0.25)$, the functional random intercepts $f_{\mu 3}\left(t, i\right)= \bcx_{\mu 3} \bm{\beta}_{\mu 3}$, and the global intercept and covariate effect $f_{\mu 4}(\bm{x}_{\mu i}) = 0.5$ and $ f_{\mu 5}(\bm{x}_{\mu i}) = 0.6 \sin(x_{2i})$ with $x_{2i} \sim unif(-3, 3)$. We simulate the functional random intercepts flexibly by P-Splines where we draw the true vector of spline-coefficients for all subjects from $\bm{\beta}_{\mu 3} \sim N(\bm{0},[(1/ \tau_s^2) \bm{\tilde{K}}_{s} + (1/ \tau_t^2)  \bm{\tilde{K}}_t ]^{-1})$ as in \eqref{eq:aniso_pen} with $\bck_t = \bcd_2^\top \bcd_2$, $\tau_s^2=1$ and $\tau_t^2=0.2$. The hazard $h_i(t)$ for every subject is calculated according to \eqref{eq:hazard} using the true survival predictor functions $\eta_{\lambda}(t)= 1.4\log((t+10)/1000)$,
  $\eta_{\gamma i}=0.5\sin(x_{1i})$, with $x_{1i} \sim unif(-3, 3)$ and $\eta_{\alpha}(t)$ varying for the two simulation settings.  Based on $h_i(t)$, survival times are generated for every subject as described in \citet{bender_generating_2005} and \citet{crowther_simulating_2013}.
Every subject is censored after $\max(\mathcal{P})$ and we additionally apply uniform censoring $U(0, 1.5 \cdot \max(\mathcal{P}))$ to the survival times. In order to mimic missing measurements in the real data, $p\%$ of the remaining longitudinal data are randomly set to missing after censoring in line with the survival times. Longitudinal obervations are obtained from $\eta_{\mu i}(t)$ by adding independent errors $\epsilon_{ij} \sim N(0, 0.3^2)$ for each $t_{ij}$ in $\bt$. 

The influence of different data structures on the estimation is assessed by simulating two different data settings in each of the two simulations settings. In the smaller data setting, $a$, observations for $n_a=150$ subjects are generated at the measurements points $\mathcal{P}_a = [0, 1, \dots, 120]$ where $p_a=75$\% of the longitudinal measurements are missing and on average 108 (72 \%) events occur, compared to $n_b=300$ subjects at the time points $P_b = [0, 3, \dots, 72]$ with $p_b=10$\% missings and 165 (55 \%) events in the larger data setting, $b$. 

In each data and simulation setting we draw $Q=200$ samples. To ensure convergence, we run the model estimation with 23000 samples, a burn-in of 3000 and a thinning of 20, yielding 1000 samples, as assessed in preliminary simulations. For each estimated model $q$ within a simulation setting we assess bias, mean-squared error (MSE) and frequentist coverage of the 95\% credibility intervals, defined by the 2.5th and the 97.5th percentiles of the MCMC samples for the posterior mean and the approximate normal intervals for the posterior mode. We evaluate bias, MSE and coverage both averaged over all time points and averaged per time point. For the predictors in the longitudinal model, i.e. $k \in \{\mu, \sigma\}$, the average bias in each sample $q$ is $B^q_{k} = \frac{1}{N} \sum_{i=1}^{n} \sum_{j=1}^{n_i} [\hat{\eta}^{q}_{k i}(t_{ij}) - \eta^{q}_{k i}(t_{ij})]$ where $\hat{\eta}_{ki}$ denotes the estimate. To assess the model fit over time we also evaluate the bias per timepoint $B^q_k(t) =  \frac{1}{n} \sum_{i=1}^{n} [\hat{\eta}^{q}_{k i}(t) - \eta^{q}_{k i}(t)]$ for all $t$ in $\mathcal{P}$. The computations for MSE and coverage are analoguous. For the survival predictors, i.e. $k \in \{\gamma, \lambda, \alpha \}$, the average bias is $B^q_{k} = \frac{1}{n} \sum_{i=1}^{n} [\hat{\eta}^{q}_{k i}(T_{i}) - \eta^{q}_{k i}(T_{i})]$ using evaluations at the subject's event times. The bias of the time-varying survival predictor $\eta_{\lambda}$, and for setting 2 also $\eta_{\alpha}$, is additionally evaluated at the fixed grid of time points $t$ in $\mathcal{P}$ as  $B^q_k(t) = \frac{1}{n} \sum_{i=1}^n [\hat{\eta}^{q}_{k i}(t) - \eta^{q}_{k i}(t)]$ with MSE and coverage computed accordingly. These error measures are then averaged over all $Q$ samples per setting. \\

For the comparison with the joint model implementation in \texttt{JMbayes} in settings 1a and 1b, data is generated with $\eta_{\alpha}(t)=1$ as time-constant. In our implementation we model the longitudinal submodel by P-splines with cubic B-splines, a second order difference penalty and 12 knots (4 internal knots), for both the overall mean as well as the individual trajectories. After application of the constraints this yields $7 \cdot n$ basis functions. For the time-varying effect of the baseline hazard, $\eta_{\lambda}$, as well as the nonlinear effect in $\eta_{\gamma}$ we use 10 knots (2 internal knots) resulting in 5 basis functions per effect after application of the constraints. In order to achieve a comparable model in the package \texttt{JMbayes} we model nonlinear effects in the longitudinal submodel and survival covariate effects by B-splines and determine the number of knots to minimize the DIC in preliminary simulations. Details on the inclusion of nonlinear effects in both submodels can be found in the source code of the ancillary material. As a result we model the longitudinal part by cubic B-splines for both the fixed and random effects with 1 internal knot for the larger data setting and without internal knots for the smaller data setting, resulting in 4 and 3 basis functions for both the fixed and random effects of time, respectively. As prior simulations had shown convergence issues when the covariance matrix of the random effects was estimated as unrestricted, we restrict it to be  diagonal, resulting in independent random effects. Also based on DIC from preliminary simulations we specify the effect in $\eta_{\gamma}$ in the survival part with cubic B-splines with 3 internal knots using 5 basis functions. We model the baseline hazard with P-splines using the default settings from \texttt{JMbayes}, i.e. a cubic B-spline basis with 17 basis functions and a second order difference penalty. For the MCMC procedure we also use the default settings of 20000 iterations, including a burn-in of 3000 and a thinning such that 2000 samples are kept.

In our second simulation, i.e. settings 2a and 2b, we specify the longitudinal trajectories as before but generate data using a time-varying association predictor  $\eta_{\alpha}(t) = \cos((t-33)/33)$ for data in $a$ and $\eta_{\alpha}(t) = \cos((t-20)/20)$ for $b$ in order to achieve a similar shape despite a differing time scale. We fit the model using the same specification as in setting 1. Additionally $\eta_{\alpha}$ is modeled as a P-spline with 10 knots (2 internal knots) resulting in 5 basis functions after application of the constraints.

\subsection{Simulation results}

The focus of the first simulation is the comparison with the package \texttt{JMbayes} regarding the accuracy of the modeling of the longitudinal trajectories and the time-constant association parameter $\eta_{\alpha}$ in settings 1a and 1b.  Table \ref{tbl:sim1} shows the MSE, bias and coverage for the estimation of $\eta_{\alpha}$. 

\begin{table}[h]
\begin{threeparttable}
\begin{center}
\caption{Posterior mean simulation results from \texttt{bamlss} and results from \texttt{JMbayes} from setting 1 (time-constant $\eta_{\alpha}$) for small ($a$) and large ($b$) data sets.}
\label{tbl:sim1}
\begin{tabular}{llcccccc}
\hline
&  & \multicolumn{2}{c}{MSE} & \multicolumn{2}{c}{bias} & \multicolumn{2}{c}{coverage} \\ 
& & a & b & a & b & a & \multicolumn{1}{c}{b} \\ 
\hline
$\eta_{\alpha}$ & \texttt{bamlss}  & $\phantom{-}0.032$ & $\phantom{-}0.016$ & $\phantom{-}0.003$ & $-0.009$ & $\phantom{-}0.925$ & $\phantom{-}0.970$ \\

& \texttt{JMbayes}  & $\phantom{-}0.049$ & $\phantom{-}0.021$ & $\phantom{-}0.100$ & $\phantom{-}0.048$ & $\phantom{-}0.840$ & $\phantom{-}0.890$ \\

$\eta_{\gamma}+\eta_{\lambda}$ & \texttt{bamlss}  & $\phantom{-}0.127$ & $\phantom{-}0.077$ & $-0.007$ & $\phantom{-}0.011$ & $\phantom{-}0.935$ & $\phantom{-}0.946$ \\

& \texttt{JMbayes}  & $\phantom{-}0.155$ & $\phantom{-}0.101$ & $-0.095$ & $-0.048$ & $\phantom{-}0.743$ & $\phantom{-}0.742$ \\

$\eta_{\mu}$ & \texttt{bamlss}  & $\phantom{-}0.022$ & $\phantom{-}0.028$ & $\phantom{-}0.001$ & $\phantom{-}0.000$ & $\phantom{-}0.944$ & $\phantom{-}0.942$ \\

& \texttt{JMbayes}  & $\phantom{-}0.031$ & $\phantom{-}0.029$ & $-0.001$ & $\phantom{-}0.008$ & $\ast$ & $\ast$\\

$\eta_{\sigma}$ & \texttt{bamlss}  & $\phantom{-}0.001$ & $\phantom{-}0.001$ & $\phantom{-}0.009$ & $\phantom{-}0.014$ & $\phantom{-}0.940$ & $\phantom{-}0.875$  \\

& \texttt{JMbayes}  & $\phantom{-}0.007$ & $\phantom{-}0.002$ & $\phantom{-}0.080$ & $\phantom{-}0.039$ & $\ast$ & $\ast$ \\

\hline
\end{tabular}
    \begin{tablenotes}
      \small
      \item $\ast$ No credibilty intervals and thus no coverage could be calculated for these predictors. 
    \end{tablenotes}
\end{center}
\end{threeparttable}
\end{table}

For both methods $\eta_{\alpha}$ is estimated more precisely and with a higher coverage in the larger data setting $b$ compared to $a$. In both data settings \texttt{bamlss} achieves lower MSE, less bias and a higher coverage in the estimation of the association compared to \texttt{JMbayes}. For \texttt{JMbayes} the coverage for $\eta_{\alpha}$ is not satisfactory in both settings (0.840 and 0.890). 
The further survival predictors, $\eta_{\gamma}$ and $\eta_{\lambda}$, are parameterized differently in the two estimation methods  with regard to the intercept term and sum-to-zero constraints. Therefore we assess only the prediction quality of $\eta_{\lambda} + \eta_{\gamma}$. We observe that \texttt{JMbayes} shows a higher bias in the estimation of the sum of these two predictors.

\begin{figure}[h]
\begin{center}
\includegraphics[width=1\textwidth]{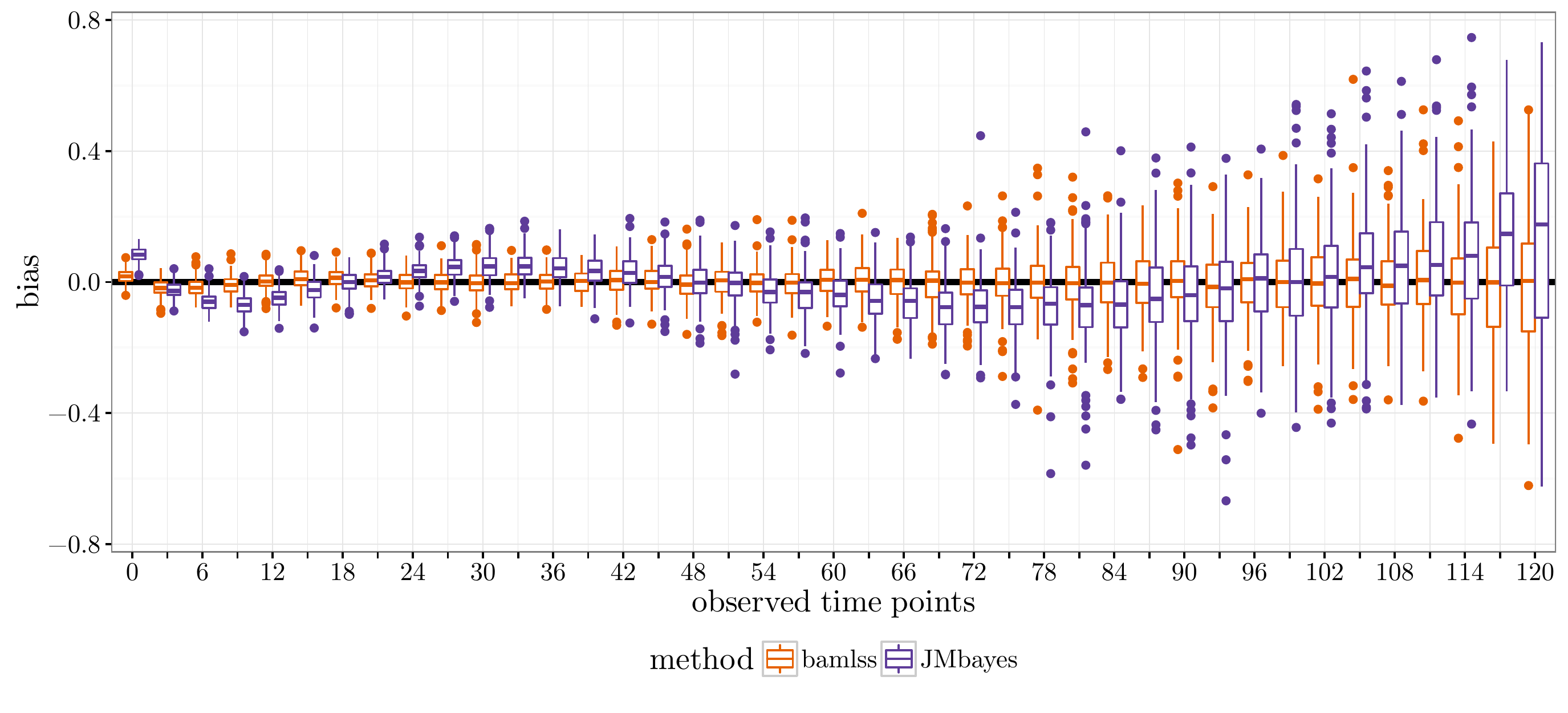}
\caption{Comparison of the bias over time for $\eta_{\mu}(t)$ estimates from \texttt{bamlss} and \texttt{JMbayes} in setting 1a.}
\label{fig:sim1_mu}
\end{center}
\end{figure}

Regarding the longitudinal submodel for $\eta_{\mu}$ both methods are fairly equal regarding the average MSE over the larger data setting (\texttt{bamlss}: 0.028 vs. \texttt{JMbayes}: 0.029), but our approach seems to be more precise in the smaller data setting (\texttt{bamlss}: 0.022 vs. \texttt{JMbayes}: 0.031). To further understand the cause of this difference we look at the bias in the estimation of $\eta_{\mu}$ over the whole observed time course for the smaller data setting. As shown in Figure \ref{fig:sim1_mu}, \texttt{JMbayes} seems to underestimate some nonlinearity of the true predictor. Both methods show higher uncertainty for later time points when, due to censoring and the occurrence of events, less information is available.

 Finally, the estimation of the error variance is more precise in \texttt{bamlss}. For the longitudinal predictors we did not achieve to calculate credibility intervals in \texttt{JMbayes}.

There are large runtime differences where \texttt{JMbayes} models took on average 4 minutes and 7 minutes for data setting $a$ and $b$, respectively, and the  implementation \texttt{bamlss}, due to the more flexible functional random effects specification, took on average 6 hours and 39 hours to run on a single core of a 2.60 GHz Intel Xeon Processor E5-2650. Through parallel computating, e.g. on 10 cores of a Linux system, the run times would reduce to 1.3 and 8.5 hours, respectively. \\

The aim of the second simulation setting is to shed light on the precision of the estimation of all predictors in the model with a special focus on the estimation of $\eta_{\alpha}$, which is nonlinear in time. Additionally, we also compare the precision of the posterior mode to the posterior mean estimation. Table \ref{tbl:sim2} gives an overview of the estimation precision of all predictors.

\begin{table}[h]
\begin{center}
\caption{Posterior mode and posterior mean simulation results for setting 2 (time-varying $\eta_{\alpha}(t)$) for small $(a)$ and large $(b)$ data sets.}
\label{tbl:sim2}
\begin{tabular}{llcccccc}
\hline
& & \multicolumn{2}{c}{MSE} & \multicolumn{2}{c}{bias} & \multicolumn{2}{c}{coverage} \\ 
& & a & b & a & b & a & \multicolumn{1}{c}{b} \\ 
\hline
$\eta_{\alpha}$ & mean  & $\phantom{-}0.172$ & $\phantom{-}0.078$ & $\phantom{-}0.007$ & $\phantom{-}0.002$ & $\phantom{-}0.940$ & $\phantom{-}0.961$  \\
  & mode  & $\phantom{-}0.177$ & $\phantom{-}0.117$ & $\phantom{-}0.058$ & $\phantom{-}0.069$ & $\phantom{-}0.608$ & $\phantom{-}0.593$ \\
$\eta_{\gamma}$ & mean  & $\phantom{-}0.097$ & $\phantom{-}0.062$ & $-0.035$ & $-0.032$ & $\phantom{-}0.931$ & $\phantom{-}0.948$ \\
 & mode & $\phantom{-}0.089$ & $\phantom{-}0.059$ & $\phantom{-}0.022$ & $-0.001$ & $\phantom{-}0.804$ & $\phantom{-}0.795$ \\
$\eta_{\lambda}$ & mean  & $\phantom{-}0.083$ & $\phantom{-}0.065$ & $\phantom{-}0.000$ & $\phantom{-}0.000$ & $\phantom{-}0.945$ & $\phantom{-}0.957$ \\
 & mode  & $\phantom{-}0.101$ & $\phantom{-}0.082$ & $\phantom{-}0.000$ & $\phantom{-}0.000$ & $\phantom{-}0.592$ & $\phantom{-}0.549$ \\
$\eta_{\mu}$ & mean  & $\phantom{-}0.022$ & $\phantom{-}0.028$ & $\phantom{-}0.000$ & $\phantom{-}0.000$ & $\phantom{-}0.943$ & $\phantom{-}0.942$ \\
 & mode  & $\phantom{-}0.025$ & $\phantom{-}0.031$ & $\phantom{-}0.000$ & $\phantom{-}0.000$ & $\phantom{-}0.882$ & $\phantom{-}0.865$ \\
 $\eta_{\sigma}$ & mean  & $\phantom{-}0.000$ & $\phantom{-}0.001$ & $\phantom{-}0.009$ & $\phantom{-}0.015$ & $\phantom{-}0.905$ & $\phantom{-}0.855$ \\
 & mode  & $\phantom{-}0.004$ & $\phantom{-}0.004$ & $-0.057$ & $-0.057$ & $\phantom{-}0.175$ & $\phantom{-}0.045$ \\
\hline 
\end{tabular}
\end{center}
\end{table}

Similarly to setting 1 we observe an effect of sample size: All survival predictors ($\eta_{\lambda}, \eta_{\gamma}, \eta_{\alpha}$) show a smaller MSE for data setting $b$ compared to $a$ probably due to the higher number of events. In contrast, the MSE is smaller for the estimation of $\eta_{\mu}$ in data setting $a$ compared to $b$ potentially due to the longer follow-up and a slightly higher number of longitudinal observations per subject.
Whereas the precision of the point estimates is overall similar or only slightly worse for the posterior mode compared to the posterior mean estimation, the coverage is not acceptable for the posterior mode but close to 95\% for the posterior mean. The only exception is the estimation of $\eta_{\sigma}$, where the coverage is somewhat lower for the posterior mean. As $\eta_{\sigma}$ is very precisely estimated and formal inference is usually not of interest for this predictor, we do not rate this under-coverage as too problematic. 

\begin{figure}[h]
\begin{center}
\includegraphics[width=1\textwidth]{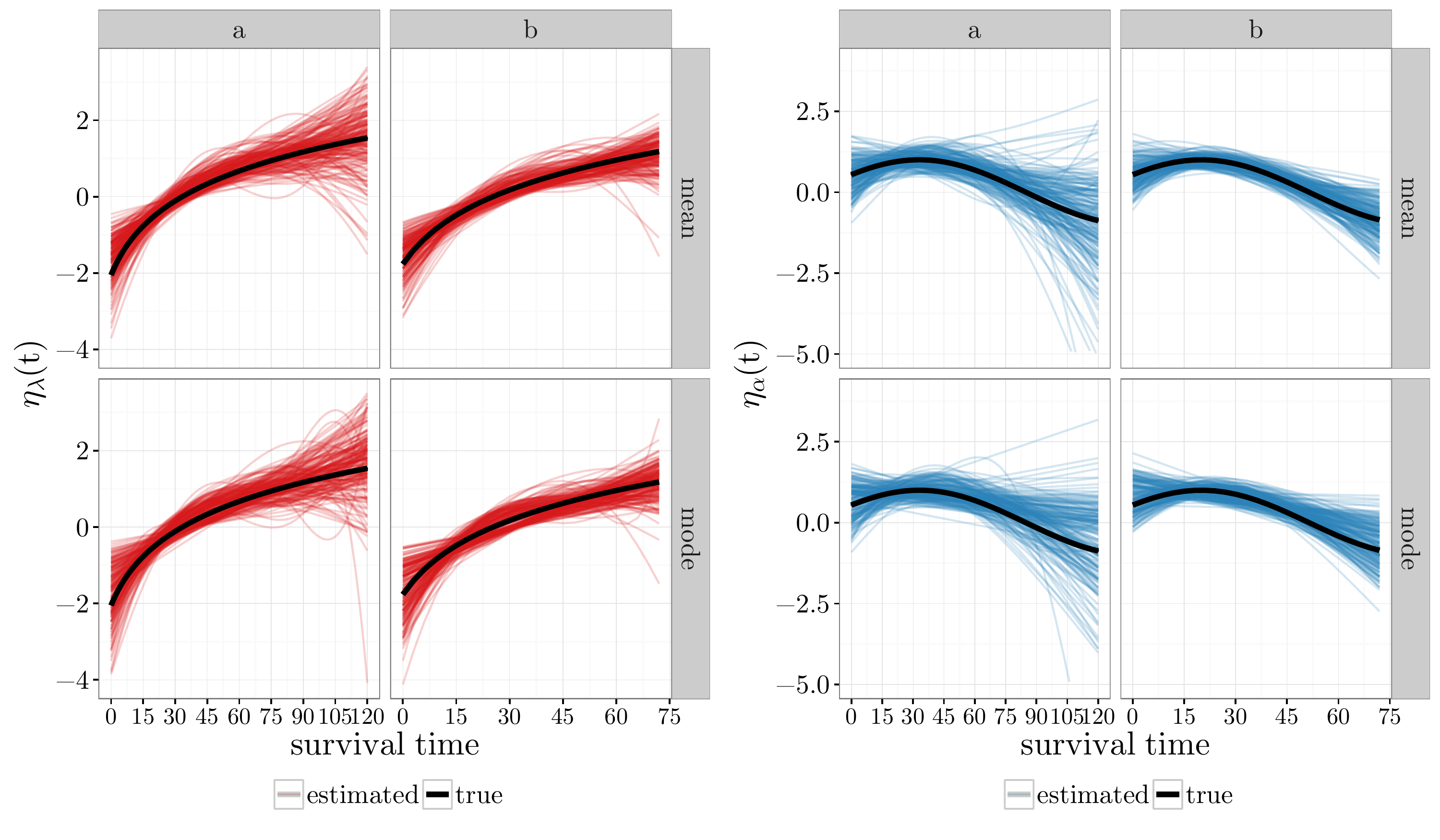}
\caption{True (black) and estimated (coloured) predictors from posterior mean und posterior mode estimation  for small ($a$) and large ($b$) data sets in simulation setting 2. Left:  $\eta_{\lambda}(t)$; right: $\eta_{\alpha}(t)$.}
\label{fig:sim2_alpha_lambda}
\end{center}
\end{figure}

In order to illustrate the precision in the time-varying effect estimates and to assess the cause of differences in MSE, Figure \ref{fig:sim2_alpha_lambda} displays the true and estimated predictors $\eta_{\lambda}(t)$ and $\eta_{\alpha}(t)$. Overall the estimated predictors match the true functions quite well. For the smaller data sets there is more uncertainty in the estimation, especially at later time points, when less subjects are still observed. 

With on average only 15 and 22 minutes to run for data setting $a$ and $b$ respectively, the posterior mode estimation has clear advantages in computation time over the more precise posterior mean estimation with 7 and 43 hours on average in this setting. Again through parallel computating on 10 cores, the time for the posterior mean estimation would reduce to 1.5 and 9.3 hours, respectively.\\

In conclusion, our simulations show that the estimation of models with constant associations between marker and event performs well, even outperforming the implementation in \texttt{JMbayes} in some aspects. The estimation of more flexible models that are newly covered by our approach in contrast to existing implementations, i.e. with a time-varying assocation parameter and the specification of flexible trajectories, is equally satisfactory. 
While the more precise posterior mean estimation is time-consuming, the posterior mode offers a computationally efficient way to quickly assess the point estimates in a given model specification, even though credibility bands are only approximate.

\section{Application}
\label{sec5}

In order to gain insights into our motivating research question we apply the model to a combined data set of two ongoing German T1D risk cohorts to investigate whether longitudinal trajectories of insulin autoantibodies (IAA) are associated with the rate of progression to T1D. Whereas different autoantibodies are diagnostic for a preclinical stage of the disease, our focus lies on the analysis of the levels of IAA as a marker from the time when it first exceeded a specific threshold, called seroconversion, to the onset of T1D or loss to follow-up. The marker IAA is most often the first autoantibody to appear \citep{babydiab_1993, babydiab_1999, hummel_brief_2004}. Both its initial value at seroconversion as well as its mean over time have been shown to be positively associated with the emergence of T1D and negatively related to the age at T1D diagnosis \citep{Steck2011, steck_predictors_2015}.

The BABYDIAB and BABYDIET studies, both propective birth cohorts with a joint study protocol, aim to investigate the natural history of T1D development. In these studies children with familial increased risk of T1D were followed from birth to the development of T1D or loss to follow-up for up to 21 years \citep{babydiab_1993, babydiab_1999, babydiab_2004, babydiet_2011}. In both studies, autoantibody measurements were taken at age 9 months and 2, 5, 8, 11, 14 and 17 years and additionally every 6 months after positive islet autoantibodies had emerged. The exact age at the emergence of clinical T1D was assessed also between study visits.

In our joint model we use data of $n = 127$ children who developed IAA during follow-up of which 69 (54\%)  progressed to T1D. The subject's progression times are censored at 15 years after seroconversion due to the extremely low sample size at later time points. In total $N = 894$ longitudinal measurements of IAA after seroconversion were used  and log-transformed $\log(IAA +1)$ for the analysis. We model subject's transformed autoantibody levels using functional random intercepts and two further covariates. First, the age at seroconversion is included as a linear effect and second a binary variable indicates whether the autoantibody was among the first autoantibodies to appear. We model the association between marker and event, $\eta_{\alpha}(t)$, to be a non-linear function of time. Further we allow the covariates in the longitudinal model to also influence the survival process directly and expect a positive association between the age at seroconversion and the time to T1D \citep{Steck2011, ziegler_seroconversion_2013}.
In our Bayesian model estimation we sample for 33000 iterations with a burnin of 3000 and thinning of 30 to obtain 1000 samples, with starting values for the posterior mean estimation obtained from the posterior mode estimates. Convergence is assessed by the inspection of traceplots, of which a subset is presented in Appendix B. In order to assess the sensitivity of the results to the number of knots we specify three models with differing numbers of knots. We specify two models using either 12 (i.e. 4 internal) knots or 20 (i.e. 8 internal) knots for the overall mean as well as the individual trajectories in the functional random intercepts and 10 (i.e. 2 internal) knots in the survival submodel. Additionally we specify a model with 20 (i.e. 8 internal) knots for nonlinear terms in both, the longitudinal and the survival submodels.

The results from the three specified models in our sensitivity analysis are highly similar for all predictors with regard to mean estimates and the credibility intervals. However we observe lower DIC for the models with more knots in the functional random intercepts along with a closer fit of the individual trajectories and more narrow credibility intervals for the estimated association $\eta_{\alpha}(t)$ (cf. Figure \ref{fig:supp_app} in Appendix B). Using more knots in the survival submodel results in a better mixing in the traceplots but a slightly higher DIC. Hence we assume the results to be robust regarding the exact number of knots and present results of the model with the lowest DIC in the following. 

As shown in Figure \ref{fig:long_est} for 5 randomly selected subjects, we are able to closely approximate the individual non-linear trajectories of IAA. The association between the marker and the onset of clinical T1D is estimated as stable over time with an average slope of -0.01 [95\% credibility interval: -0.08, 0.06]. 

\begin{figure}[h] 
\begin{center}
   \includegraphics[width=0.5\textwidth]{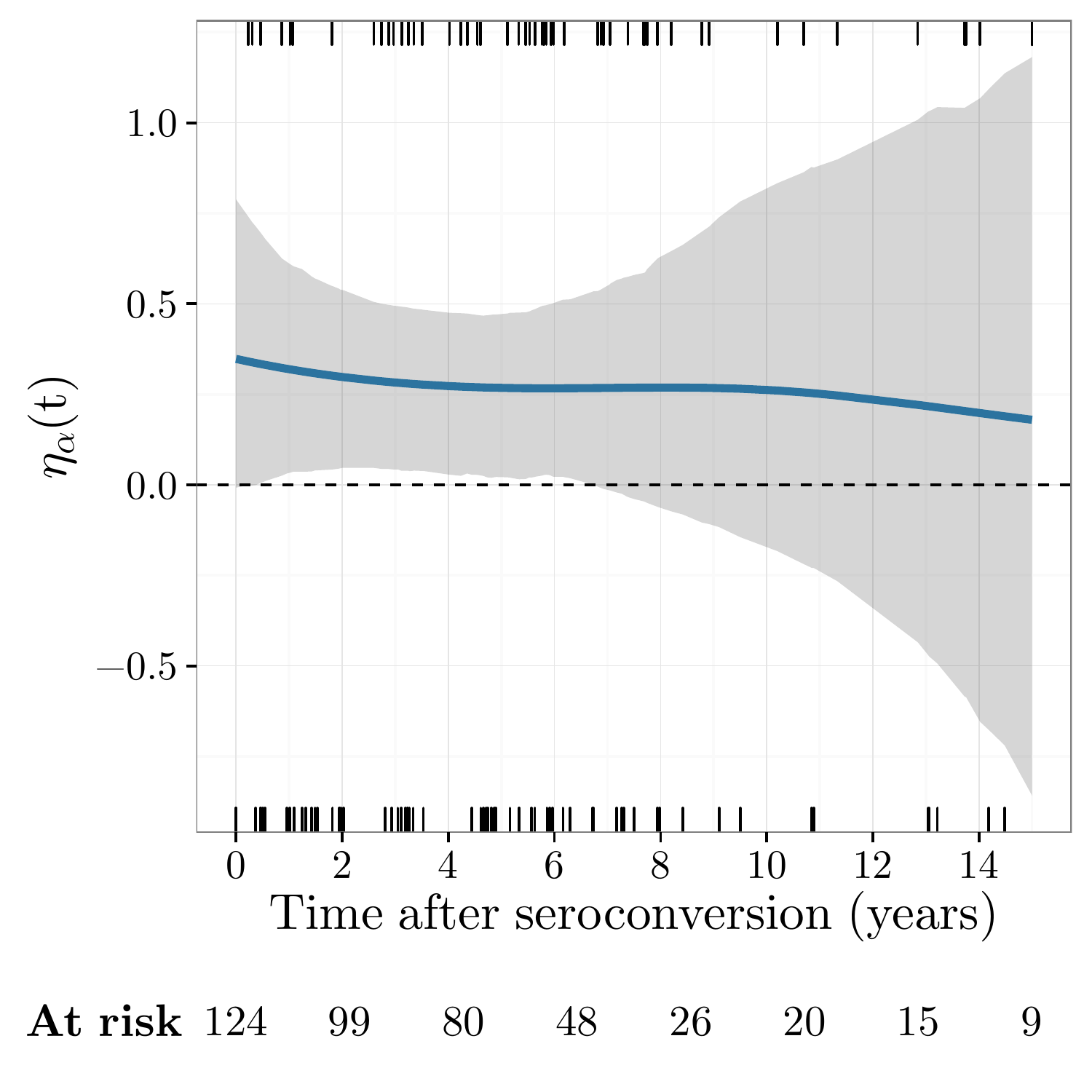} 
  \caption{Estimated posterior mean of $\eta_{\alpha}(t)$ with 95\% pointwise credibility bands (shaded area), observed event times (rugs bottom) and censoring times (rugs top), and number of subjects at risk per time point (bottom).} 
\label{fig:app_effects}
\end{center}
\end{figure}

The average slope was defined as the mean over the first derivative of the association $\eta_{\alpha}(t)$ evaluated at all observed event and follow-up times $\bct$, and its posterior distribution can be easily obtained by numerically deriving $\eta^{\prime}_{\alpha}(t)$ in every sample. The credibility interval for the estimated association is above 0 from 0.5 to 6.5 years after seroconversion and there is more uncertainty when less information is available, i.e. when less event and follow-up times are observed and when less subjects remain in the risk set, as indicated by the credibility intervals (Figure \ref{fig:app_effects}). In the longitudinal submodel we observe that trajectories have a lower level, if subjects seroconverted at an older age (in years, $\beta_{\mu 4}= -0.07$; 95\% credibility interval: [-0.14, -0.01]) and a higher level if IAA was amongst the first markers to appear ($\beta_{\mu 5}= 0.90$; [0.24, 1.55]). In the survival submodel the log-hazard is decreased if IAA was amongst the first markers to appear ($\beta_{\gamma 2}= -0.94$; [-1.69, -0.14]). In sum if IAA is amongst the first markers to appear the log-hazard is reduced by 0.71. This net effect can be derived as the sum of the direct effect in $\eta_{\gamma}$ and the indirect effect in $\eta_{\alpha} \cdot \eta_{\mu}$ with an average association of $\eta_{\alpha}=\frac{1}{n} \sum_{i} \eta_{\alpha}(T_i) = 0.25$. Additionally we do not observe a direct effect of the age at seroconversion ($\beta_{\gamma 3}= -0.09$; [-0.20, 0.01]). 

In line with previous findings \citep{Steck2011, steck_predictors_2015} these results indicate that the quantitative levels of the marker IAA are informative for the rate of progression to T1D in the first years after seroconversion with higher levels increasing the hazard of T1D. The direct relationship between the hazard and the baseline covariate age at seroconversion is not supported by the model, suggesting that the previously established influence of this covariate on T1D progression may be mediated by the marker levels, i.e. the effect in the respective log-hazard is reduced if the marker levels over time are taken into account as in our flexible parameterization. 
We do not observe a time-varying association between IAA and the hazard of T1D over time. There is much uncertainty around the nonlinear time-varying estimate of the association $\eta_{\alpha}(t)$, potentially as a result of the flexibility in the estimation in combination with the amount of data in the survival part.

\section{Discussion and Outlook}
\label{sec6}
We presented a flexible joint model that allows to fit a broad range of joint model specifications using structured additive predictors for all model components. The approach is fully implemented in the R-package \texttt{bamlss}. While the framework is very flexible, as illustrated by Table \ref{tbl:effects}, the focus in this work lies on the flexible modeling of individual trajectories and the specification of a time-varying association between marker and event.

The proposed model shows satisfactory performance in various simulation settings and has the potential to offer new insights into complex relationships between biomarkers and time-to-event processes. Our methods development was motivated by a specific research question from T1D studies and two corresponding data sets. We saw that even by combining the two cohorts, the sample size of the data set considered in the application in Section \ref{sec5} is at the lower limit for the complexity of our model, as indicated by our simulation study and by the width of the credibility intervals in the applied results. Nevertheless we found a positive association between a disease-related biomarker and the occurrence of clinical T1D. Although our model allows for a time-varying association between the biomarker and the event process at least in this small data set it was estimated to be roughly constant. In consequence our flexible model can also be used to check the modeling assumptions of simpler models that are commonly used. We aim to further explore the relationship between T1D-specific autoantibodies and the progression to T1D in a larger data set from a different, multinational T1D cohort (with sample size exceeding data setting $b$ in our simulations) as in \citet{steck_predictors_2015}.

Due to the complexity of the model and its estimation, the computation speed is still a drawback in our implementation. Hence we are constantly working on speeding up the computations further. As shown in simulation 2, the posterior mode estimation offers a computationally efficient way to obtain point estimates from a flexible joint model before starting the full MCMC sampling. These posterior mode estimates show a precision similar to that of the posterior mean estimates. However, the credibility intervals obtained from posterior modes are not wide enough, potentially due to the fact that the uncertainty around the variance parameters $\tau_{km}^2$ is not included in the credibility intervals. In consequence, only the credibility intervals of the posterior mean estimates should be used for inference. 

As is well known in the survival context, the number of potential parameters in the model is limited by the number of observed events \citep{HarrellLeeMark1996}. This also holds in our approach for the predictors in the survival part of the model, $\eta_{\lambda}$, $\eta_{\gamma}$, and $\eta_{\alpha}$. We achieve to alleviate this issue to some extent by the penalized approach, which decreases the effective number of degrees of freedom and thus allows for a richer model than would be possible without a penalty. Still, we recommend to model only those effects as non-linear functions, where a strong indication for non-linearity is given. 

Within the framework of the presented additive joint model several further extensions are possible. As a next step we aim to extend the model by including the derivative of the longitudinal trajectories to model the event process similar to \cite{Ye2008}, \cite{brown_assessing_2009} and \cite{rizopoulos_combining_2014}, allowing to model the potentially time-varying  association between changes in the marker and the hazard. Further, functional historical effects of the trajectories, including information on the history of the marker \citep{Malfait2003, Gellar2014}, could potentially offer additional insights into complex relationships between markers and event processes.

\begin{acknowledgement}
We thank Lorenz Lachmann, Claudia Matzke, Joanna Stock, Stephanie Krause, Annette Knopff, Florian Haupt, Maren Pfl\"{u}ger, Marlon Scholz and Anita Gavrisan (all: Institute for Diabetes Research, Helmholtz Zentrum M\"{u}nchen) for data collection and expert technical assistance, Ramona Puff (Institute for Diabetes Research, Helmholtz Zentrum M\"{u}nchen) for laboratory management, and Peter Achenbach (Institute for Diabetes Research, Helmholtz Zentrum M\"{u}nchen) and Ezio Bonifacio (Center for Regenerative Therapies Dresden and Paul Langerhans Institute Dresden, Technische Universit\"{a}t Dresden) for overseeing antibody measurement and for fruitful discussions and advice on modeling T1D-specific autoantibodies. We also thank all pediatricians and family doctors in Germany for participating in the BABYDIAB Study. Furthermore we thank Fabian Scheipl (Ludwig-Maximilians-Universit\"at M\"unchen) for advice on modeling functional random intercepts. 
The work was supported by the JDRF (JDRF-2-SRA-2015-13-Q-R) and by grants from the German Federal Ministry of Education and Research (BMBF) to the German Center for Diabetes Research (DZD e.V.). Meike K\"{o}hler’s work was supported by a grant from the Helmholtz International Research Group (HIRG-0018) and Sonja Greven acknowledges funding from the German research foundation (DFG) through Emmy Noether grant GR 3793/1-1.

\end{acknowledgement}
\vspace*{1pc}

\noindent {\bf{Conflict of Interest}}

\noindent {\it{The authors have declared no conflict of interest.}}

\newpage

\appendix
\section*{Appendix A}
\label{sec8}

We derive score vectors and Hessians for the regression coefficients of every predictor. We introduce some further notation to formulate these derivatives.
For the time-varying predictors of the survival part $k \in \{\lambda, \alpha, \mu\}$ the design matrix $\bcx_k(\bct)$ denotes the $n \times p_k$ matrix of evaluations at the vector of survival times $\bct$. For the time-varying predictors of the longitudinal part $k \in \{\mu, \sigma\}$ the $N \times p_k$ design matrix $\bcx_k(\bt)$ contains the evaluations at all observed subject-specific timepoints $\bt$. 
Let $\ell$ denote the log-likelihood, i.e. the sum of the contributions of the longitudinal and survival submodels defined in \eqref{eq:l_surv} and \eqref{eq:l_long}. In more detail, the full likelihood is 
\begin{align*}
\ell\left[\bm{\theta} | \bct, \bm{\delta}, \bm{y}\right] = &  \bm{\delta}^{\top}\left[\bcx_{\lambda}(\bct)\bm{\beta}_{\lambda} + \bcx_{\gamma}\bm{\beta}_{\gamma} + \bcx_{\alpha}(\bct)\bm{\beta}_{\alpha} \cdot \bcx_{\mu}(\bct)\bm{\beta}_{\mu} \right] \\
& -\sum_{i=1}^{n}\exp\left(\bm{x}_{\gamma i}^{\top}\bm{\beta}_{\gamma}\right)\int_{0}^{T_{i}}\exp\left[\bm{x}_{\lambda i}^{\top}\left(u\right)\bm{\beta}_{\lambda}+\bm{x}_{\alpha i}^{\top}\left(u\right)\bm{\beta}_{\alpha}\left(\bm{x}_{\mu i}^{\top}\left(u\right)\bm{\beta}_{\mu}\right)\right] \ du \\ 
& -\frac{N}{2}\log(2\pi)- \bm{1}_{N}^{\top} \bcx_{\sigma}\left(\bt\right)\bm{\beta}_{\sigma} -\frac{1}{2}(\bm{y}-\bcx_{\mu}\left(\bt\right)\bm{\beta}_{\mu})^{\top} \bm{R}^{-1}(\bm{y}-\bcx_{\mu}\left(\bt\right)\bm{\beta}_{\mu})
\end{align*}

\subsection*{Score Vectors}
\label{Scores}
\begin{align*}
\bm{s}(\bm{\beta}_{\mu}) = \frac{\partial\ell}{\partial\bm{\beta}_{\mu}} 
 = &  \bcx_{\mu}\left(\bt\right)^{\top}\bm{R}^{-1}\left(\bm{y}-\bcx_{\mu}\left(\bt\right)\bm{\beta}_{\mu}\right)+ \bcx_{\mu}^{\top}\left(\bct\right)\diag(\bm{\delta})\left[\bcx_{\alpha}\left(\bct\right)\bm{\beta}_{\alpha} \right]\\
    & -\sum_{i=1}^{n}\exp\left(\bm{x}_{\gamma i}^{\top}\bm{\beta}_{\gamma}\right)\int_{0}^{T_{i}}\omega_i(u) \  \bm{x}_{\alpha i}^{\top}\left(u\right)\bm{\beta}_{\alpha}\bm{x}_{\mu i}\left(u\right)du\\
\bm{s}(\bm{\beta}_{\gamma})  = \frac{\partial\ell}{\partial\bm{\beta}_{\gamma}} 
 = & \bm{\delta}^{\top}\bcx_{\gamma}-\sum_{i=1}^{n}\exp\left(\bm{x}_{\gamma i}^{\top}\bm{\beta}_{\gamma}\right)\bm{x}_{\gamma i}\int_{0}^{T_{i}}\omega_i(u) \ du\\
\bm{s}(\bm{\beta}_{\alpha}) = \frac{\partial\ell}{\partial\bm{\beta}_{\alpha}} 
 = & \bcx_{\alpha}^{\top}\left(\bct\right)\diag(\bm{\delta})\left[\bcx_{\mu}\left(\bct\right)\bm{\beta}_{\mu}\right] -\sum_{i=1}^{n}\exp\left(\bm{x}_{\gamma i}^{\top}\bm{\beta}_{\gamma}\right)\int_{0}^{T_{i}}\omega_i(u) \ \bm{x}_{\alpha i}\left(u\right)\left(\bm{x}_{\mu i}^{\top}\left(u\right)\bm{\beta}_{\mu}\right)du\\
\bm{s}(\bm{\beta}_{\lambda}) = \frac{\partial\ell}{\partial\bm{\beta}_{\lambda}} 
 = & \bm{\delta}^{\top}\bcx_{\lambda}\left(\bct\right)-\sum_{i=1}^{n}\exp\left(\bm{x}_{\gamma i}^{\top}\bm{\beta}_{\gamma}\right)\int_{0}^{T_{i}}\omega_i(u) \ \bm{x}_{\lambda i}\left(u\right)du\\
\bm{s}(\bm{\beta}_{\sigma}) = \frac{\partial\ell}{\partial\bm{\beta}_{\sigma}} 
 = & - \bcx_{\sigma}\left(\bt\right)^{\top} \bm{1}_{N} + \left[\bcx_{\sigma}\left(\bt\right) \odot \left(\bm{y}-\bcx_{\mu}\left(\bt\right)\bm{\beta}_{\mu}\right)\right]^{\top}
\bm{R}^{-1}\left(\bm{y}-\bcx_{\mu}\left(\bt\right)\bm{\beta}_{\mu}\right)\\
\end{align*}

\subsection*{Hessian}
\begin{align*}
\bm{H}(\bm{\beta}_{\mu}) =\frac{\partial^{2}\ell}{\partial\bm{\beta}_{\mu}\partial\bm{\beta}_{\mu\top}} 
 = & 
-\bcx_{\mu}\left(\bt\right)^{\top}\bm{R}^{-1}\bcx_{\mu}\left(\bt\right) \\
& -\sum_{i=1}^{n}\exp\left(\bm{x}_{\gamma i}^{\top}\bm{\beta}_{\gamma}\right)\int_{0}^{T_{i}}\omega_i(u) \ \left[\bm{x}_{\alpha i}^{\top}\left(u\right)\bm{\beta}_{\alpha}\right]^{2}\bm{x}_{\mu i}\left(u\right)\bm{x}_{\mu i}^{\top}\left(u\right)du\\
\bm{H}(\bm{\beta}_{\gamma})  = \frac{\partial^{2}\ell}{\partial\bm{\beta}_{\gamma}\partial\bm{\beta}_{\gamma}^{\top}} 
 = & 
-\sum_{i=1}^{n}\exp\left(\bm{x}_{\gamma i}^{\top}\bm{\beta}_{\gamma}\right)\bm{x}_{\gamma i}\bm{x}_{\gamma i}^{\top}\int_{0}^{T_{i}}\omega_i(u) \ du\\
\bm{H}(\bm{\beta}_{\alpha}) = \frac{\partial^{2}\ell}{\partial\bm{\beta}_{\alpha}\partial\bm{\beta}_{\alpha}^{\top}} 
 = & -\sum_{i=1}^{n}\exp\left(\bm{x}_{\gamma i}^{\top}\bm{\beta}_{\gamma}\right)\int_{0}^{T_{i}}\omega_i(u) \ \left[\bm{x}_{\mu i}^{\top}\left(u\right)\bm{\beta}_{\mu}\right]^{2}\bm{x}_{\alpha i}\left(u\right)\bm{x}_{\alpha i}^{\top}\left(u\right)du\\
\bm{H}(\bm{\beta}_{\lambda}) = \frac{\partial^{2}\ell_{i}}{\partial\bm{\beta}_{\lambda}\partial\bm{\beta}_{\lambda}^{\top}} 
 = & -\sum_{i=1}^{n}\exp\left(\bm{x}_{\gamma i}^{\top}\bm{\beta}_{\gamma}\right)\int_{0}^{T_{i}}\omega_i(u) \ \bm{x}_{\lambda i}\left(u\right)\bm{x}_{\lambda i}^{\top}\left(u\right)du\\
\bm{H}(\bm{\beta}_{\sigma}) = \frac{\partial^{2}\ell}{\partial\bm{\beta}_{\sigma}\partial\bm{\beta}_{\sigma}^{\top} } 
 = & -2 \left[\bcx_{\sigma}\left(\bt\right) \odot \left(\bm{y}-\bcx_{\mu}\left(\bt\right)\bm{\beta}_{\mu}\right)\right]^{\top}
\bm{R}^{-1}\left[\bcx_{\sigma}\left(\bt\right) \odot \left(\bm{y}-\bcx_{\mu}\left(\bt\right)\bm{\beta}_{\mu}\right)\right]\\
\end{align*}
where  $\omega_i(u) = \exp\left[\bm{x}_{\lambda i}^{\top}\left(u\right)\bm{\beta}_{\lambda}+\bm{x}_{\alpha i}^{\top}\left(u\right)\bm{\beta}_{\alpha}\left(\bm{x}_{\mu i}^{\top}\left(u\right)\bm{\beta}_{\mu}\right)\right]$ and  $\bm{R}=\diag\left(\exp\left[\bcx_{\sigma}\left(\bt\right)\bm{\beta}_{\sigma}\right]^2\right)$.

\newpage

\section*{Appendix B}
\label{sec9}
\setcounter{figure}{0}
\renewcommand{\thefigure}{A\arabic{figure}}  

\begin{figure}[h] 
  \subfloat[\label{fig:supp_app_long}]{%
    \includegraphics[width=0.45\textwidth]{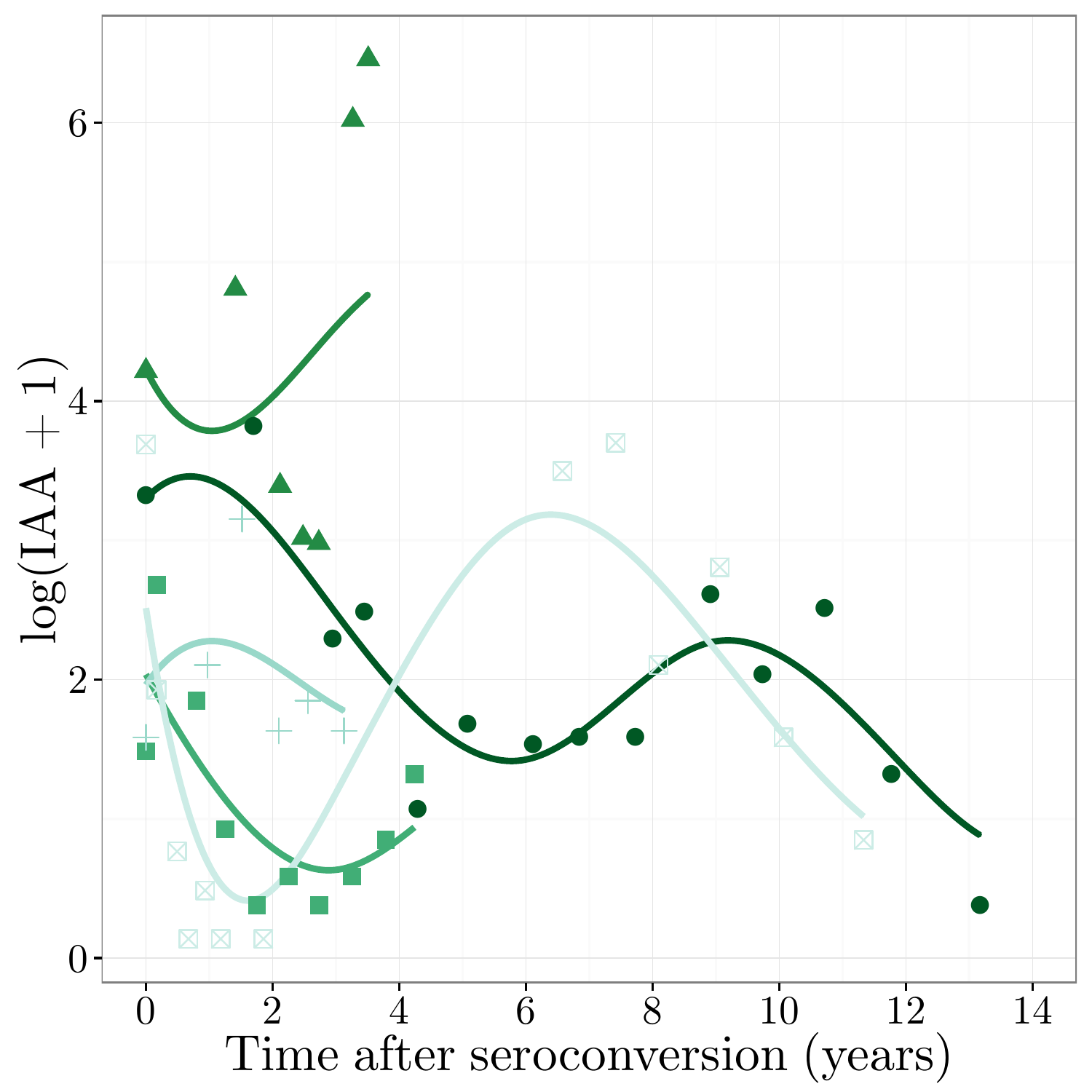} 
  } 
  \subfloat[\label{fig:supp_app_alpha}]{%
    \includegraphics[width=0.45\textwidth]{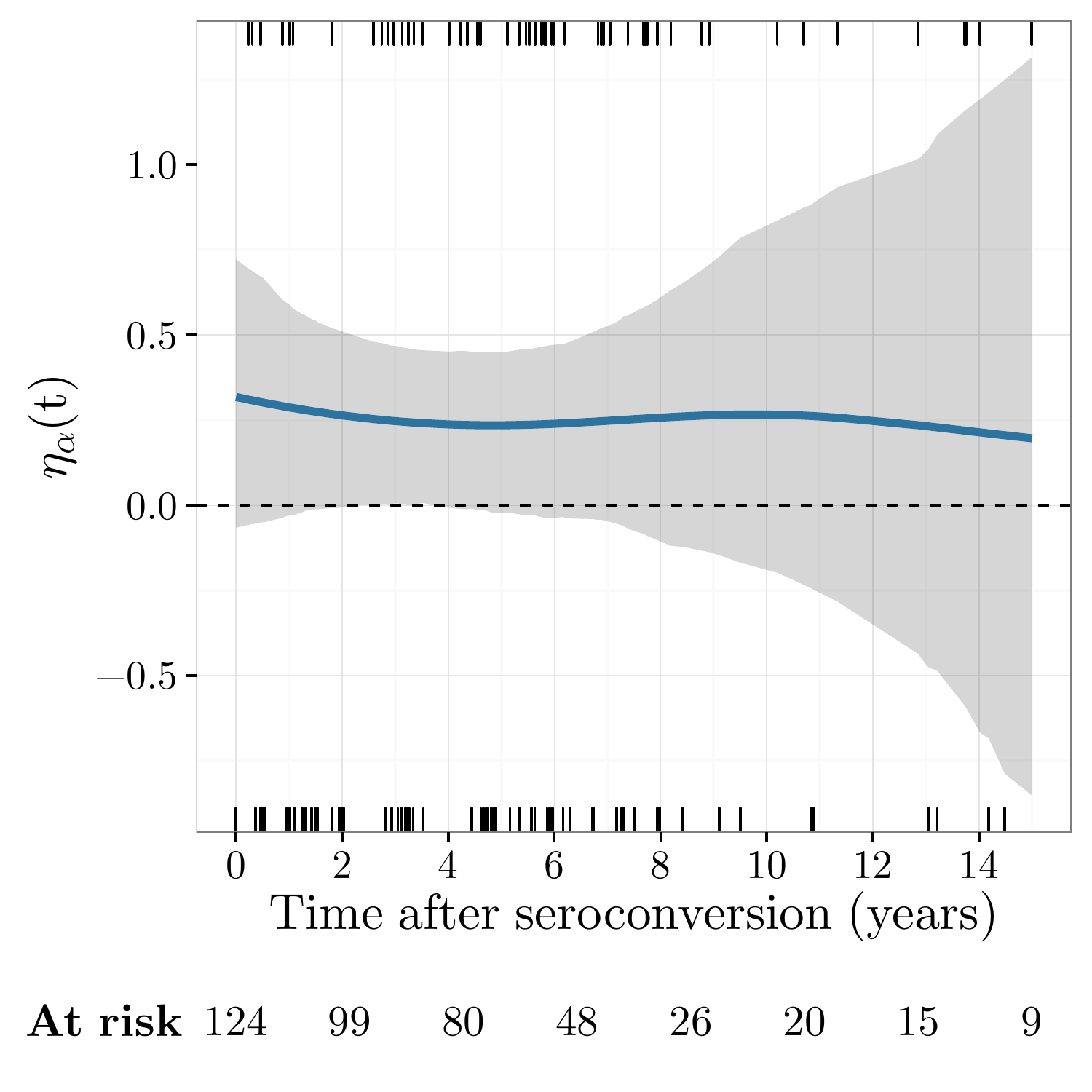} 
  } 
  \caption{Results from the sensitivity analysis for the T1D data using 12 (i.e. 4 internal) knots for both the overall mean and the functional random intercepts in the longitudinal submodel. (a) Observed values (points) and estimated trajectories (lines) of the longitudinal marker values of $\log(IAA + 1)$ for five randomly selected subjects; (b) Estimated posterior mean of $\eta_\alpha(t)$ with 95\% pointwise credibility bands (shaded area), observed event times (rugs bottom) and censoring times (rugs top), and number of subjects at risk per time point (bottom). } 
\label{fig:supp_app}
\end{figure}

\begin{figure}[h] 
    \includegraphics[width=0.9\textwidth]{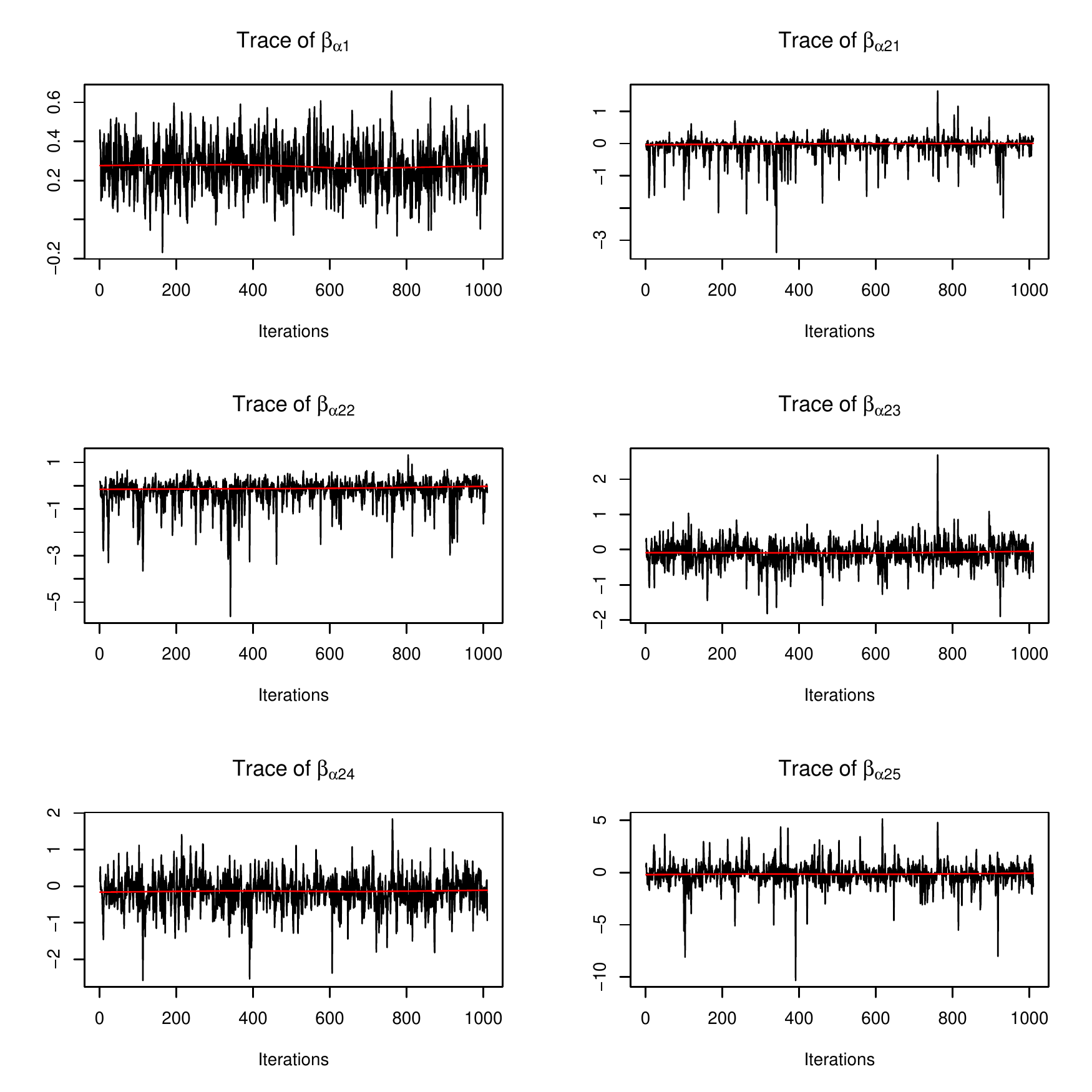} 
    \caption{Traceplots of the posterior samples for the intercept $\beta_{\alpha 1}$ and the coefficient vector $\bm{\beta}_{\alpha 2}$ in $\eta_{\alpha}(t)$.} 
    \label{fig:supp_trace_mu}
\end{figure}

\begin{figure}[!htbp] 
    \includegraphics[width=0.8\textwidth]{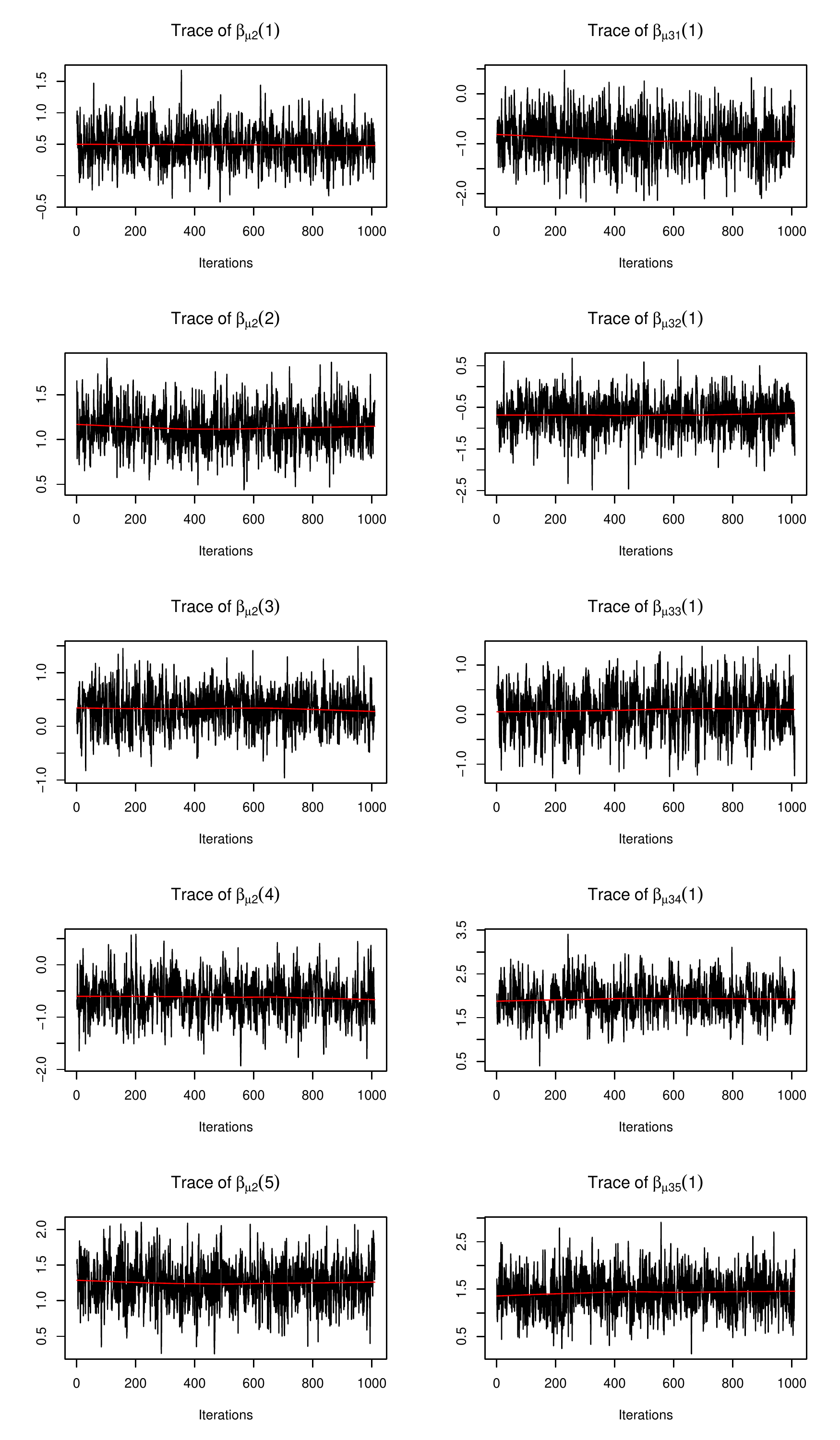} 
    \caption{Traceplots of the posterior samples for the random intercepts $\beta_{\mu 2}(i)$ of subjects $i = 1, \dots, 5$, and the coefficient vector ${\bm\beta}_{\mu 3}(t, i)$ for subject $i =1$ in $\eta_\mu(t)$. } 
    \label{fig:supp_trace_mu}
\end{figure}

\clearpage

\bibliographystyle{biometrical}
\bibliography{paper-1}

\end{document}